\documentclass[pdflatex,sn-mathphys-num]{sn-jnl}% Math and Physical Sciences Numbered Reference Style

\usepackage{amssymb}

\usepackage{nccmath}
\usepackage{url,hyperref}
\usepackage{graphicx}
\usepackage{subfig}
\usepackage{tikz}
\usepackage{mathtools}
\usepackage{amsmath}
\usepackage{blkarray}
\usepackage{rotating} 
\usepackage{esvect}
\usepackage{enumerate}
\usepackage{algorithmicx}
\usepackage{algorithm}
\usepackage{algpseudocode}
\usepackage{color}
\usepackage{listings}%
\usepackage{mathrsfs}
\usepackage{textcomp}%
\usepackage{manyfoot}%
\usepackage{booktabs}%
\usepackage{multirow,bbm}
\usepackage{lineno,balance}
\usepackage[utf8]{inputenc}
\usepackage{makecell}
\usepackage{pifont}%
\definecolor{mypink3}{cmyk}{0, 0.47808, 0.10429, 0.012}
\usepackage{amssymb}
\usepackage{amsmath}
\usepackage{multicol}
\newcommand{\caseScopeNode}{\textit{Single-Node}}
\newcommand{\caseScopeMulti}{\textit{Multi-Node}}
\newcommand{\caseScopeSystem}{\textit{Full System}}
\newcommand{\caseScopeApp}{\textit{Full App}}

\newcommand{\casePhaseOperational}{\textit{Operational}}
\newcommand{\casePhaseExperimental}{\textit{Experimental}}

\newcommand{\caseAcquisitionStandardTelemetry}{\textit{Standard Telemetry}}
\newcommand{\caseAcquisitionCustomInstrumentation}{\textit{Custom Instrumentation}}
\newcommand{\caseAcquisitionExperimentalInstrumentation}{\textit{Experimental Instrumentation}}

\begin{document}
\title{A Taxonomy of Performance Metrics for the Distributed Computing Continuum}
% \title{\huge Performance Evaluation of the Distributed Computing Continuum}
\author*[1]{\fnm{Praveen Kumar} \sur{Donta}}\email{praveen@dsv.su.se}

\author[2]{\fnm{Boris} \sur{Sedlak}}\email{b.sedlak@dsg.tuwien.ac.at}

\author[1]{\fnm{Alfreds} \sur{Lapkovskis}}\email{alfreds.lapkovskis@dsv.su.se}

\author[3]{\fnm{Alaa} \sur{Saleh}}\email{alaa.saleh@helsinki.fi}

\author[4]{\fnm{Ying} \sur{Li}}\email{liying2@cse.neu.edu.cn}

\author[5]{\fnm{Victor} \sur{Casamayor~Pujol}}\email{victor.casamayor@upf.edu}
\author[6]{\fnm{Ilir} \sur{Murturi}}\email{ilir.murturi@uni-pr.edu}
\author[7]{\fnm{Manuel} \sur{Otero Barbasan}}\email{motero6@us.es}
\author[2,8]{\fnm{Schahram} \sur{Dustdar}}\email{schahram.dustdar@upf.edu}

\affil*[1]{\orgdiv{Department of Computer and Systems Sciences}, \orgname{Stockholm University}, \orgaddress{\street{NOD Borgarfjordsgatan 12}, \city{Stockholm}, \postcode{106 91}, \state{Stockholm}, \country{Sweden}}}

\affil[2]{\orgdiv{Distributed Systems Group}, \orgname{TU Wien}, \orgaddress{\street{Argentinierstrasse 8 / 194-02}, \city{Vienna}, \postcode{1040}, \state{Wien}, \country{Austria}}}

\affil[3]{\orgdiv{Department of Computer Science} , \orgname{University of Helsinki}, \orgaddress{\street{Fabianinkatu 33}, \city{Helsinki}, \postcode{00014}, \state{Helsinki}, \country{Finland}}}

\affil[4]{\orgdiv{School of Computer Science and Engineering
}, \orgname{Northeastern University}, \orgaddress{\street{No. 195 Chuangxin Road}, \city{Shenyang}, \postcode{110169}, \state{Liaoning}, \country{China}}}

\affil[5]{\orgdiv{Department of Engineering}, \orgname{Universitat Pompeu Fabra}, \orgaddress{\street{Roc Boronat, 138}, \city{Barcelona}, \postcode{08018}, \state{Barcelona}, \country{Spain}}}

\affil[6]{\orgdiv{Department of Mechatronics}, \orgname{University of Prishtina}, \orgaddress{\street{Rr. George Bush, Nr.31}, \city{Prishtina}, \postcode{10000}, \state{Prishtina}, \country{Kosovo}}}

\affil[7]{\orgdiv{Department of Computer Languages and Systems}, \orgname{Universidad de Sevilla}, \orgaddress{\street{C. San Fernando 4}, \city{Sevilla}, \postcode{41004}, \state{Sevilla}, \country{Spain}}}

\affil[8]{\orgname{ICREA, Barcelona},  \orgaddress{\street{Passeig Lluís Companys 23}, \city{Barcelona}, \postcode{08010}, \state{Barcelona}, \country{Spain}}}

\abstract{Performance evaluation is essential for understanding, comparing, and improving computing systems, including Distributed Computing Continuum Systems (DCCS). In recent years, computational requirements have changed substantially with the growth of artificial intelligence and large-scale data-driven applications. These application tasks are increasingly distributed between resource-intensive data centers and resource-constrained edge environments. In this context, novel computing continuum architectures and algorithms are emerging, creating a need for transparent and consistent performance evaluation. However, existing evaluation practices often focus on isolated dimensions, such as computation, networking, energy efficiency, or application-level quality, and therefore provide only a partial view of cross-layer DCCS behavior. This paper presents a structured taxonomy of performance metrics for DCCS. The taxonomy organizes metrics into computing-level, network-level, and application/user-level categories, while also highlighting emerging dimensions such as sustainability, observability, adaptability, data locality, migration awareness, and continuum fragmentation. Further, we provide mathematical formulations and discuss their relevance to heterogeneous and dynamic continuum environments. We also summarize metric acquisition requirements in terms of acquisition scope, acquisition phase, and measurement method. These requirements help clarify whether a metric can be collected from a single node, multiple nodes, or the full system, and whether it is more suitable for operational monitoring or experimental evaluation. }

\maketitle

\section{Introduction}\label{sec:Introduction}
%This paragrpah discuss about the introduction to DCCS, charcteristics, etc. 
Distributed computing continuum systems (DCCS) exemplify a profound shift in the computing paradigm, ushering in an era where traditional boundaries blur and dynamic synergies emerge \cite{casamayor2022distributed, donta2023exploring}. This paradigm envisions a seamless continuum of computing resources that includes data centers, cloud infrastructures, edge computing nodes, and IoT devices. It allows applications and services to intelligently assign tasks, data, and services across the computing continuum, unlike a uniform computing model \cite{pujol2023edge}. Traditionally, data centers provided massive processing and storage capabilities, while cloud platforms extended this reach and provided scalable, on-demand resources \cite{buyya2018manifesto}. The edge computing environment facilitates low-latency processing by putting data sources close to the data processors, minimizing data transit times, and improving applications' response times \cite{hua2023edge, dustdar2020towards, dustdar2021towards}. In addition to this continuum, a growing IoT system introduces distributed sensors and actuators. These sensors generate huge amounts of real-time data that can be analyzed and responded to accordingly \cite{li2015internet}. In this computing paradigm, applications take advantage of the unique strengths of each computing node and dynamically allocate workloads along the computing continuum. For example, limited data and time-critical applications like autonomous vehicles or healthcare flourish at the edge.
In contrast, data-intensive analytics such as artificial intelligence (AI), Machine learning (ML), and Large language models (LLMs) leverage the computational power of cloud data centers \cite{dustdar2022distributed}. This way maximum benefit is achieved by maximizing efficiency and resource usage, reducing operating costs, and energy consumption \cite{casamayor2023fundamental}. However, measuring the efficiency and performance of DCCS is challenging due to the vast variety of heterogeneous resources. 

Distributed Computing Continuum Systems (DCCS) represent a shift from centralized computing toward a heterogeneous and distributed execution model that spans cloud data centers, fog nodes, edge servers, mobile devices, and Internet of Things (IoT) devices \cite{casamayor2022distributed, donta2023exploring}. The DCCS paradigm enables applications to place computation, data, and services across multiple layers of the continuum according to application requirements and resource availability \cite{pujol2023edge}. Traditionally, data centers and cloud platforms provided large-scale processing, storage, and on-demand resource provisioning \cite{buyya2018manifesto}. More recently, edge and fog computing have extended this model by placing computation closer to data sources, thereby reducing data transmission delay and supporting latency-sensitive applications \cite{hua2023edge, dustdar2020towards, dustdar2021towards}. At the same time, the IoT has introduced large numbers of distributed sensors and actuators that continuously generate real-time data and require timely processing and response \cite{li2015internet}.

In this continuum-based model, applications can exploit the strengths of different computing layers. Latency-sensitive and context-aware applications, such as autonomous vehicles, healthcare monitoring, industrial automation, and robotics, can benefit from processing near the edge. In contrast, computation-intensive and data-intensive workloads, such as artificial intelligence (AI), machine learning (ML), and large language models (LLMs), often require the high processing capacity and storage resources of cloud data centers \cite{dustdar2022distributed}. It is noteworthy that, based on model size the inference are also distributed among edge. By dynamically distributing workloads across the continuum, DCCS can improve resource utilization, reduce communication overhead, lower operational cost, and support energy-aware execution \cite{casamayor2023fundamental}. Recently, several architectural advances and algorithmic strategies have emerged for DCCS, including intelligent orchestration~\cite{rawlley2025artificial,gong2026intelligent,fan2025vehicular,shahid2026iot}, adaptive task offloading~\cite{kumar2026task,mohajer2026joint,11346001}, resource-aware scheduling~\cite{salimi2026dynamic,ayouni2026resource}, service migration~\cite{hua2025intelligent,liu2025service,du2025online,zeng2025towards,bozkaya2025optimizing}, edge intelligence, and AI-assisted management~\cite{10925537,11120402,saleh2026usercentrixagenticmemoryaugmentedai,ye2026nesyedgeneurosymbolictrustworthyselfhealing,saleh2025memindex}. As these approaches make different assumptions about resources, workloads, mobility, and communication conditions, their performance must be evaluated transparently and consistently to support fair comparison and reproducible conclusions~\cite{11270701}. Therefore, transparent and consistent performance evaluation is increasingly needed. However, existing evaluation practices often focus on isolated dimensions, such as computational performance, network efficiency, energy consumption, reliability, or application-level quality. While these metrics are useful, they do not always capture the cross-layer behavior of DCCS. 

Performance evaluation is essential for understanding, designing, and improving computing systems, and this importance also extends to DCCS \cite{11270701}. It supports informed decisions about system architecture, resource provisioning, deployment configuration, and cost-performance trade-offs. It also helps system designers identify bottlenecks, compare alternative configurations, and optimize existing systems according to application requirements \cite{sedlak_markov_2024}. In this sense, performance evaluation provides practical insight into how hardware, software, communication mechanisms, and operational procedures affect the overall behavior of a computing system \cite{kant1992introduction}. However, evaluating DCCS requires a broader perspective than evaluating conventional centralized or homogeneous distributed systems because DCCS consist of distributed, interconnected, and heterogeneous computing nodes spanning cloud, fog, edge, mobile, IoT, and sensor layers. These nodes differ in processing capacity, memory, storage, energy availability, mobility, network connectivity, and failure behavior. Therefore, DCCS performance cannot be fully captured by a single metric or by metrics designed for only one system layer. Instead, benchmarking and performance measurement methods should be selected according to the application domain, deployment scenario, and evaluation objective. %In DCCS, performance can be evaluated from multiple complementary perspectives, including computing-level, network-level, and application/user-level behavior. Computing-level metrics capture resource usage and task execution efficiency, network-level metrics describe communication quality, connectivity, and data delivery performance, and application/user-level metrics reflect service quality, task success, user experience, and application-specific requirements. Figure~\ref{fig:taxonomy} summarizes these metric categories and presents the taxonomy used throughout this paper.

%In this paragraph, I am trying to discuss the 'Common goals of performance metrics for DCCS'
The goals of DCCS performance evaluation depend on the system context, application requirements, and deployment objectives~\cite{casamayor_pujol_deepslos_2024}. Nevertheless, several evaluation goals are common across DCCS environments, including performance optimization, scalability assessment, Quality of Service (QoS) assurance, bottleneck detection, energy efficiency, cost-effectiveness, reliability, security, and user experience improvement~\cite{cao_better_2023, lu_qos-aware_2024}. Performance optimization focuses on reducing latency, improving resource utilization, and distributing workloads efficiently across the continuum. Scalability assessment determines whether the system can handle increasing workloads without unacceptable performance degradation. QoS evaluation verifies whether the system satisfies application-specific requirements such as delay, throughput, availability, and service continuity~\cite{cilic_performance_2023}. Bottleneck detection helps identify performance limitations caused by network latency, bandwidth constraints, overloaded nodes, inefficient algorithms, or resource shortages. Energy and cost evaluation are also critical because DCCS combine low-power IoT devices with energy-intensive cloud and edge infrastructures, which directly affects operational cost and sustainability~\cite{herrera_active_2026}. Ultimately, performance evaluation helps improve both system efficiency and user satisfaction~\cite{lilja2005measuring}.

\begin{figure}[h]
    \centering
    \includegraphics[width=0.95\textwidth]{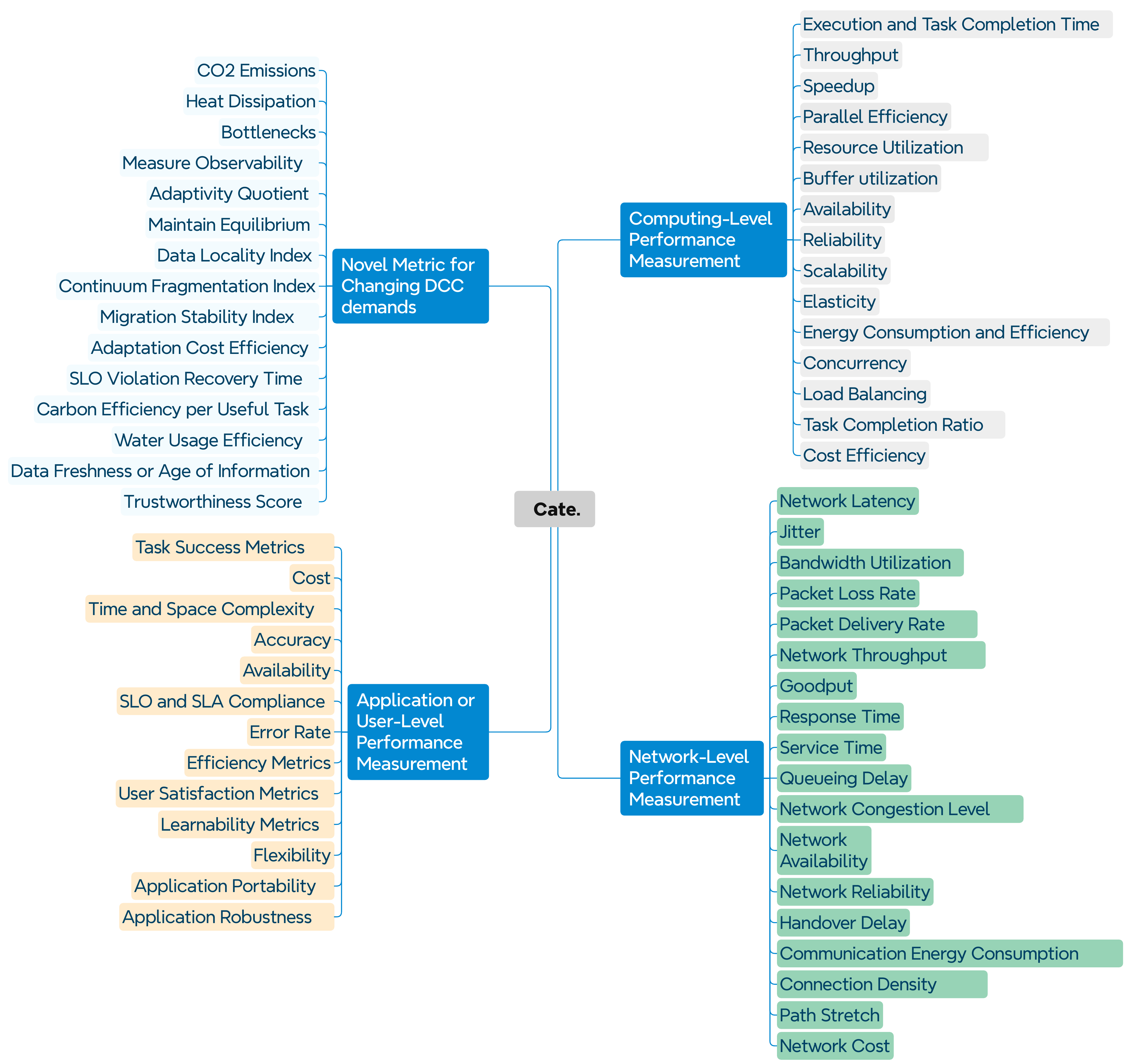}
    \caption{A Taxonomy of  Performance Metrics for the DCCS}
    \label{fig:taxonomy}
\end{figure}

While keeping all these complexities in mind, we provide various performance metrics in this paper, and a taxonomy is shown in Figure~\ref{fig:taxonomy}. In short, the contributions of this paper are summarized as follows.
\begin{itemize}
    \item We present a structured taxonomy of performance metrics for DCCS, considering the heterogeneity of cloud, fog, edge, mobile, IoT, and sensor layers.
    
    \item We classify performance metrics into computing-, network-, and application- or user-level categories to support systematic evaluation of DCCS behavior.
    
    \item We provide mathematical formulations for representative metrics, providing precise and comparable performance measurement across different computing configurations.
    
    \item We discuss metric acquisition requirements, including acquisition scope, acquisition phase, and acquisition method, to clarify how each metric can be measured in practical scenarios. We also provide a discussion on metrics selection criteria while evaluating the performance.
    
    \item We highlight novel performance metrics, including sustainability, observability, adaptability, energy awareness, and continuum fragmentation, which reflect the changing requirements of modern computing environments.
\end{itemize}
% \begin{itemize}
%     \item We provide various performance metrics according to their goals and characteristics regarding the DCCS. While designing the performance metrics, we consider DCCS complexity such as variability and heterogeneity of computing devices. 
%     \item We explore mathematical formulations for various performance metrics at the computing level, at the network level, and at the application or user level. 
%     \item To demonstrate these metrics' applicability in real-time DCCS scenarios, we present a case study for robotics. 
% \end{itemize}

The remaining sections of this paper are organized as follows. In section~\ref{sec:SystemModel}, we provide a detailed DCCS system model, and a taxonomy of performance metrics. In section~\ref{sec:Criteria}, we provide various criteria to be followed to consider particular performance metrics and feasibility analysis. 
Section~\ref{sec:ComputingLevel} discusses various computing-level performance metrics and their mathematical formulae. Section~\ref{sec:NetworkLevel} provides a variety of network-level performance metrics and mathematical formulae. In Section~\ref{sec:ApplicationLevel}, we discuss various performance metrics at application or user-level along with their mathematical formulae. Section~\ref{sec:novelmetrics} introduced several novel performance metrics need for the changing computational demands. 
Finally, we conclude our paper with a future scope in Section~\ref{sec:Conclusion}.

%%%% Comment, if any information missing in Introduction or possible changes.
%%%% am I aligned? 

\section{System Model}\label{sec:SystemModel}
\begin{figure}[t]
    \centering
    \includegraphics[width=0.75\textwidth]{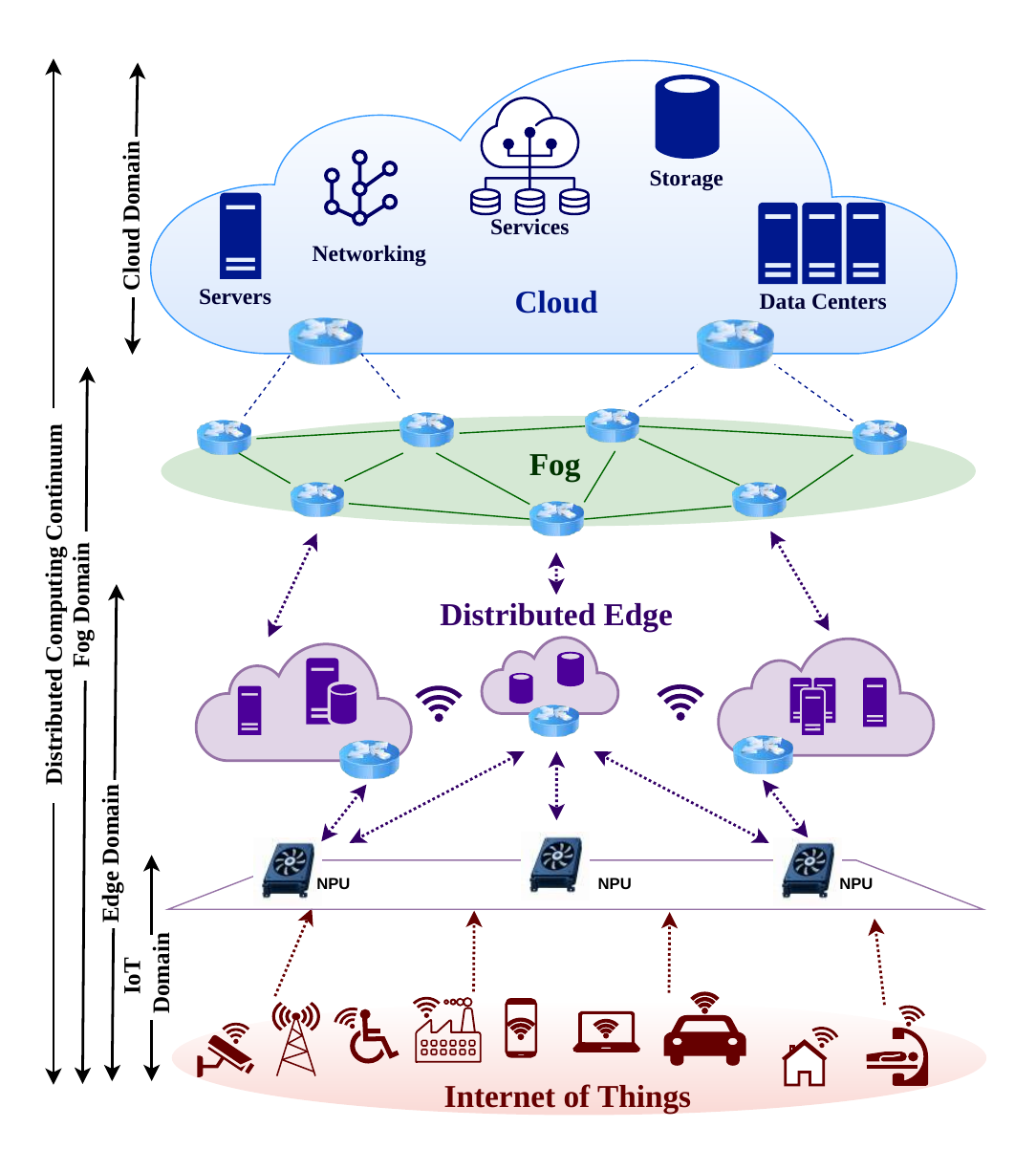}
    \caption{The general architectural view of Distributed Computing Continuum and its components \cite{11270701}}
    \label{fig:SystemModel}
\end{figure}

This section provides a detailed system model of the computing continuum through mathematical notations. In addition, a graphical representation is also provided in Figure~\ref{fig:SystemModel}. The most frequently used notations are summarized using Table~\ref{tab:notations}. 

\begin{table*}[!b]
\centering
\caption{Frequently used notations.}
\label{tab:notations}
% \footnotesize
\resizebox{\columnwidth}{!}{%
\begin{tabular}{ll|ll}
\hline
\textbf{Notation} & \textbf{Meaning}  &
\textbf{Notation} & \textbf{Meaning} \\
\hline

\(\mathcal{S}\) & DCCS system & 
\(n\) & Total number of devices in \(\mathcal{S}\) \\

\(\mathcal{C}\) & Set of cloud/data centers &
\(c\) & Number of cloud/data centers, \(c=|\mathcal{C}|\)  \\

\(\mathcal{F}\) & Set of fog nodes & 
\(f\) & Number of fog nodes, \(f=|\mathcal{F}|\)  \\

\(\mathcal{E}\) & Set of edge nodes & 
\(e\) & Number of edge nodes, \(e=|\mathcal{E}|\)  \\

\(\mathcal{M}\) & Set of mobile devices & 
\(m\) & Number of mobile devices, \(m=|\mathcal{M}|\)  \\

\(\Gamma\) & Set of IoT devices & 
\(\iota\) & Number of IoT devices, \(\iota=|\Gamma|\)  \\

\(\Psi\) & Set of sensor nodes & 
\(\varsigma\) & Number of sensor nodes, \(\varsigma=|\Psi|\)  \\

\(\mathcal{N}\) & Set of computing-capable nodes & 
\(\zeta\) & Number of computing-capable nodes, \(\zeta=|\mathcal{N}|\) \\

\(\mathcal{N}^{all}\) & Set of all devices in \(\mathcal{S}\)  &
\(\rho\) & Number of data-producing nodes \\

\(\mathcal{V}_i\) & Set of VMs in cloud/data center \(C_i\)  &
\(k_i\) & Number of VMs in \(C_i\), \(k_i=|\mathcal{V}_i|\)\\

\(V_{ir}\) & VM \(r\) in cloud/data center \(C_i\) & 
$E_i = 0$ & Operational state of node \(j\)  \\

$\mathbb{C}_{ij}$ & Direct connectivity indicator between nodes \(i\) and \(j\) & 
\(\mathcal{L}\) & Set of communication links \\

\(\mathcal{T}\) & Set of tasks & 
\(q\) & Number of generated tasks, \(q=|\mathcal{T}|\)\\

\(T_i\) & Task \(i\) & 
\(\mathcal{T}^{c}\) & Set of completed tasks  \\

\(\mathcal{T}^{f}\) & Set of failed tasks & 
\(\mathcal{T}^{c}_{\delta}\) & Set of tasks completed within deadlines \\

\(x_{ij}\) & Task assignment indicator for task \(i\) on node \(j\)&
\(\omega_i\) & Computational workload of task \(i\)  \\

\(d_i\) & Input data size of task \(i\) & 
\(r_i\) & Output/result size of task \(i\)  \\

\(\tau_i\) & Arrival time of task \(i\) & 
\(\delta_i\) & Deadline of task \(i\)  \\

\(\mu_j\) & Processing capacity of node \(j\)  &
\(B_{ij}\) & Available bandwidth between nodes \(i\) and \(j\)  \\

\(\Delta t\) & Observation window & 
\([t_0,t_1]\) & Start and end times of \(\Delta t\) \\

\hline
\end{tabular}}
\end{table*}

DCCS ($\mathcal{S}$) consists of \(n\) devices, including a set of cloud/data centers
\(\mathcal{C}=\{C_1,C_2,\ldots,C_c\}\), where \(c=|\mathcal{C}|\), a set of fog nodes
\(\mathcal{F}=\{F_1,F_2,\ldots,F_f\}\), where \(f=|\mathcal{F}|\), a set of edge nodes
\(\mathcal{E}=\{E_1,E_2,\ldots,E_e\}\), where \(e=|\mathcal{E}|\), a set of mobile devices
\(\mathcal{M}=\{M_1,M_2,\ldots,M_m\}\), where \(m=|\mathcal{M}|\), a set of IoT devices
\(\Gamma=\{I_1,I_2,\ldots,I_{\iota}\}\), where \(\iota=|\Gamma|\), and a set of sensor nodes
\(\Psi=\{\psi_1,\psi_2,\ldots,\psi_{\varsigma}\}\), where \(\varsigma=|\Psi|\).
%
% \(\Psi=\{\psi_1,\psi_2,\ldots,\psi_{\varsigma}\}\), where \(\varsigma=|\Psi|\).
%
We assume that $\Psi$ are data producers and they are not capable of computing any data. So, they transmit their data to the nearest base station, i.e., $\Gamma$, $\mathcal{M}$, or $\mathcal{E}$ using point-to-point communication mechanism through Bluetooth Low Energy (BLE), Zigbee, Cellular, or Wi-Fi communication channels \cite{cheikhrouhou2016secure}. Usually, $\Psi$ are operated using a limited powered battery to supply energy \cite{kumar2019machine}. The $\Gamma$ have higher computing power than $\Psi$, and connected with other devices ($\Gamma$, $\mathcal{M}$, $\mathcal{E}$, $\mathcal{C}$, or $\mathcal{F}$) using Internet, in contrast, $\Psi$ usually uses Intranet as discussed before. These nodes are also data producers. In general, we assume the number of SNs is higher than IoT in $\mathcal{S}$ i.e., $\iota \leq \varsigma$. $\Gamma$ use either low power battery or a regular power supply to input the energy depending on the application. For example, applications like smart buildings can be supplied directly with electricity, whereas smart agriculture applications will require batteries. 

Edge devices ($\mathcal{E}$) are closer to the data producer (For example, Gateways, Routers, or Switches) and provide computing services with low latency. Depending on the application, sometimes $\Gamma$ acts as edge devices. The $\mathcal{E}$ perform various functions, including routing, transmission, processing, filtering, monitoring, translation, and storage of processed data. These are also associated with higher memory and computing powers than $\Gamma$. Due to this, the energy consumption of $\mathcal{E}$ is higher than $\Gamma$ and $\Psi$. These devices also can operate both battery or regular power supply depending on application, and location of the device deployment. Usually, DCCS use less $\mathcal{E}$ than $\Gamma$, i.e., $e \leq \iota$ to minimize overall cost of $\mathcal{S}$. Fog nodes ($\mathcal{F}$) are placed between edge and cloud or data centers to make the computation close to the data source. In some applications, fog nodes have higher memory or computing power than edge nodes, and in some other cases, both are similar \cite{hong2019resource}. $\mathcal{F}$ do not produce any data, whereas they mainly perform computing tasks and store or revert the decisions on processed data. Since the energy need of $\mathcal{F}$ is higher than $\Gamma$, connecting with a regular power supply is a suitable option. Setup $\mathcal{F}$ are more expensive than edge devices, but lower than cloud or data centers. So, in $\mathcal{S}$ maintaining $(f < e) $ or $ (f < \iota)$. $\mathcal{E}$ and $\mathcal{F}$ are stationary, and their locations are fixed in $\mathcal{S}$. 

Mobile devices ($\mathcal{M}$) are changing their geographical location timely due to their mobility feature. Most of the applications, the configurations of $\mathcal{M}$ are similar to either $\mathcal{E}$ or $\mathcal{F}$ \cite{donta2023learning}. However, $\mathcal{M}$ can be data producers in some cases, for example, mobile phones or autonomous vehicles. Since $\mathcal{M}$ are moving, it is not possible to provide a fixed electricity supply, and they are possibly operating through a battery or wireless energy transmissions \cite{wang2023wireless}. The communication with the $\Gamma$, and $\Psi$ is also complex \cite{mao2017survey}, but the services to these devices can be performed quickly with a minimum number of devices. However, the number of $\mathcal{M}$ always to be minimal compared to $\mathcal{E}$, $\Gamma$, and $\Psi$ to minimize overall cost of $\mathcal{S}$. Cloud ($\mathcal{C}$) provides on-demand computational or storage resources, and their resources are unlimited. Since $\mathcal{C}$ is unlimited resources, it is not necessary to maintain a huge number of cloud centers, i.e., $c=1$. The $c>1$, only for the case of huge amount of data (such as Generative AI \cite{jo2023promise,baidoo2023education} or Large Language Models \cite{thirunavukarasu2023large,kasneci2023chatgpt}) which is not possible to manage at one machine or store the data in multiple replications. Due to high computing power, and maintaining high availability, it is necessary to supply continuous energy through a regular power supply. Each $C_i$ further split in to $k$ virtual machines (VMs) ($\mathcal{V}_i = \{V_{i1}, V_{i2}, ...., V_{ik}\}, \forall \ (1\leq i \leq c) $).

As discussed above, all devices in $\mathcal{S}$ have no computational power (For example, $\Psi$), the computing nodes are assumed as $\zeta$, which is treated as $\zeta = (c+f+e+m+\iota)$. The total number of data producers is $\rho= (\varsigma+\iota+e+m)$, whereas $\mathcal{E}$, and $\mathcal{M}$ properties may vary from application to application. To optimize the overall $\mathcal{S}$ cost, the $\zeta$ is optimized according to ($c < f < e \leq m \leq \iota$) and ($\zeta < \rho$). Each computing node in $\mathcal{S}$ can be one of several states such as active or operational ($0$), inactive or fail ($1$). For example, $E_i = 0$ means the edge node $i \forall i\in (1,e)$ is active or operational. Similarly, $V_{ik} =1$ means the VM $k$ of cloud $C_i$ is inactive or failed. 
(\textbf{Note}: Depends on the need, these can be further split individual to know the accurate state of any computing node. This means not all active nodes are operational and not all inactive nodes are failed. In this case, each state is assigned using different numbers from 0 to 3.) It is also necessary to identify whether a node (within the same class or a different class) is able to exchange information or communicate with each other. We can maintain a connection matrix $\mathbb{C}_{ij}  = {0,1}$, where $(i,j \in n)$ and $(i\neq j)$, here $(n = c+f+e+m+\iota+\varsigma)$. For example, $\mathbb{C}_{ij}=0$ means nodes $i$ and $j$ cannot directly communicate, whereas $\mathbb{C}_{ij}=1$ means they can directly communicate without using an intermediate node. 

% \textcolor{blue}{Possible alternative for device state:}
% To identify the state of each device, and considering that there are 4 possible states, we can represent the state as a 4 dimensional vector where each dimension maps to a state, and therefore, can be True (1) or False (0). Hence, if possible states are: operational, active, inactive, and fail; the state of a device that is operational would be represent as $\Vec{E}_i = (1,0,0,0)$, while if the device is inactive will be as $\Vec{E}_i = (0,0,1,0)$

% {\color{red}There are several other notations available in $\mathcal{S}$, which can be filled during addressing subsequent sections. 

% We can also talk about over all performance in terms of QoS or Cost .. Maybe later
% }

% {\color{red}%Elaborate
\section{Metrics Selection Criteria and Feasibility Analysis}
\label{sec:Criteria}
In general, evaluating all performance metrics simultaneously is practically challenging due to monitoring overhead and system complexity. Also, a single metric is usually not sufficient to describe the behavior of the whole continuum. Therefore, when selecting metrics, system architects must consider how they affect the overall complexity: the more parts of the system a metric spans and the more experimental its nature, the more difficult it becomes to implement.  Further, the selected metrics should depend on the purpose of the evaluation. For example, an IoT-based health monitoring application may prioritize latency, reliability, availability, and energy consumption, while an industrial automation system may focus on response time, fault tolerance, synchronization accuracy, and network stability. Similarly, a large-scale smart city application may require metrics related to scalability, throughput, resource utilization, and data processing efficiency. Therefore, metric selection must be guided by the operational goals of the DCC system rather than by the availability of metrics alone. Further, we also provide metric wide feasibility within this section.

\subsection{Metrics Selection Criteria}
% DCC or IoT performance can be described using the metrics shown in section~\ref{sec3} and subsection~\ref{section41}. However, 

The first and most important criterion is \textit{relevance}. A metric should be directly related to the objectives of the DCC system under study. Since DCC environments distribute tasks across IoT, edge, fog, and cloud resources, the metric should capture the effect of this distribution on system performance. For instance, latency is relevant when the objective is to support real-time decision-making at the edge, while energy consumption is relevant when the system includes battery-powered IoT devices. Resource utilization is important when evaluating how efficiently workloads are allocated across edge and cloud nodes. A metric that does not reflect the main objective of the system may lead to incomplete or misleading conclusions.
Another essential property is \textit{sensitivity}. A suitable metric should be able to detect small but meaningful changes in system behavior. In DCC systems, performance may change because of variations in network conditions, workload intensity, node mobility, resource availability, task placement, or data volume. A sensitive metric should reflect these changes clearly. For example, if computation is moved from the cloud to an edge server, the response time metric should show whether the change reduces delay. Similarly, if the number of connected IoT devices increases, scalability and throughput metrics should reveal whether the system can still maintain acceptable performance.

The selected metric should also be \textit{consistent}, means that the metric must have the same definition, unit, and measurement method across different experiments, configurations, and DCC layers. This is particularly important because DCC systems contain heterogeneous resources with different processing capacities, communication technologies, and energy profiles. For example, comparing latency between an IoT-to-edge path and an IoT-to-cloud path is only meaningful when latency is measured in the same way in both cases. Without consistency, it becomes difficult to compare different architectures, algorithms, or deployment strategies fairly.

A performance metric must also be \textit{repeatable}, means that the metric can be measured using the same procedure whenever the experiment is repeated under the same conditions. This property is necessary to ensure that the evaluation results are reliable. In DCC environments, repeatability can be challenging because of dynamic factors such as changing network bandwidth, device mobility, workload fluctuations, and node failures. Therefore, the measurement process should be clearly defined. The experiment should specify where the metric is measured, how often it is measured, what tools are used, and which system conditions are controlled.

The metric should be \textit{easy to understand} and \textit{easy to use}. A useful metric should have a clear meaning and should be simple enough to interpret by researchers, developers, and system operators. For example, response time, packet delivery ratio, CPU utilization, and energy consumption are easier to understand than complex composite indicators with unclear definitions. Simplicity is important because DCC systems are already complex. If the metric itself is difficult to interpret, it may reduce the usefulness of the evaluation. However, simplicity should not come at the cost of accuracy. The metric should remain meaningful and should still represent the behavior of the system correctly.

A good metric should provide sufficient \textit{coverage} of the DCC behavior. Since the DCC spans multiple layers, the selected metrics should capture performance from different perspectives. These perspectives may include communication performance, computation performance, resource efficiency, reliability, scalability, and energy efficiency. For example, communication-related metrics may include latency, bandwidth usage, jitter, and packet loss. Computation-related metrics may include task execution time, processing delay, and workload distribution. Resource-related metrics may include CPU utilization, memory usage, storage usage, and network utilization. Reliability-related metrics may include failure rate, availability, and service continuity. Energy-related metrics may include device energy consumption, node lifetime, and energy per processed task.

The metric should be as \textit{independent} as possible from external influences that are not part of the system being evaluated. In practice, DCC performance may be affected by background traffic, hardware differences, environmental conditions, wireless interference, simulation parameters, or measurement noise. A good metric should minimize the impact of such external factors, or these factors should be clearly controlled and reported. For example, when comparing two task scheduling algorithms, both algorithms should be evaluated under the same workload, network conditions, and resource constraints. Otherwise, the observed difference may come from the environment rather than from the algorithm itself.

The selected metrics should also support \textit{comparability}. One of the main purposes of performance evaluation is to compare different DCC architectures, algorithms, configurations, or deployment strategies. Therefore, metrics should be chosen in a way that enables fair comparison. For example, when comparing cloud-only processing with edge-assisted processing, the evaluation may include response time, bandwidth consumption, energy usage, task completion rate, and cost. These metrics help show not only which approach performs better, but also why it performs better.
% }

\subsection{Feasibility Analysis}
\label{subsec:feasibility}
This section propose analysis aiming to provide an objective assessment of metric feasibility. Instead of classifying metrics according to subjective criteria such as whether they are easy or difficult to obtain, we characterize their feasibility through a set of data acquisition dimensions. These dimensions provide a structured way to analyze the effort required to obtain each metric, since collecting information from multiple entities or requiring additional mechanisms for data collection increases the complexity of the acquisition process. The analysis considers the following data acquisition dimensions:
\begin{enumerate}
    \item \textbf{Scope (the where):} defines the minimum source of information required to obtain a metric. The considered options are \caseScopeNode{}, \caseScopeMulti{}, \caseScopeSystem{}, and \caseScopeApp{}, depending on whether the required information can be obtained from a single computational node, multiple interacting nodes, the entire DCCS, or all components belonging to a specific application.

    \item \textbf{Phase (the when):} defines the context in which the required data is collected. The considered options are \casePhaseOperational{} and \casePhaseExperimental{}. The former refers to data collected during normal system operation, while the latter refers to data collected through controlled evaluation scenarios.

    \item \textbf{Method (the how):} Defines how the required data is collected, that is, the degree of modification or integration required in the system to enable data collection. The options are classified as follows:
    \begin{itemize}
        \item \caseAcquisitionStandardTelemetry{}: No modifications are needed; it uses the built-in, out-of-the-box telemetry mechanisms already available in the system.
        \item \caseAcquisitionCustomInstrumentation{}: Requires modifying the application by adding custom code or specific configurations to extract tailored internal data.
        \item \caseAcquisitionExperimentalInstrumentation{}: Requires integrating non-conventional components, such as external hardware sensors, unestablished frameworks, or novel measurement procedures.
    \end{itemize}
\end{enumerate}

These dimensions are used to analyze the feasibility of acquiring each metric. A broader \textbf{scope} increases the difficulty of collecting and correlating data from multiple entities. The \textbf{phase} affects feasibility because metrics requiring experimental scenarios may need controlled deployments, specific configurations, or comparisons between different executions instead of relying on normal system operation. The \textbf{method} impacts feasibility by increasing the required level of technical knowledge and the number of modifications needed to obtain the data.

\section{Computing-level Performance Measurement}\label{sec:ComputingLevel}
Computing-level performance measurement evaluates how efficiently the computing resources of a DCCS are used to execute tasks under dynamic workload, mobility, heterogeneity, and resource constraints. In the considered system model, the computing-capable devices are represented by $\zeta=(c+f+e+m+\iota)$, where cloud/data centers, fog nodes, edge nodes, mobile devices, and IoT devices are assumed to have computing capability, while sensor nodes $\Psi$ are mainly considered data producers. Therefore, the computing-level performance of $\mathcal{S}$ depends not only on the amount of available resources, but also on how these resources are used, shared, scaled, balanced, and maintained over time. Table~\ref{tab:caseComputationMetrics} summarizes the acquisition requirements associated with each metric through three dimensions: minimal acquisition \textbf{scope}, acquisition \textbf{phase}, and acquisition \textbf{method}.
\begin{table}[!t]
\centering
\caption{Acquisition requirements for computation-oriented metrics.}
\label{tab:caseComputationMetrics}
\begin{tabular}{|l|c|c|c|}
\hline
\textbf{Metric} & \textbf{Scope} & \textbf{Phase} & \textbf{Method} \\
\hline
\textbf{Execution Time} & \caseScopeNode{} & \casePhaseOperational{} & \caseAcquisitionStandardTelemetry{} \\
\hline
\textbf{Task Completion Time} & \caseScopeNode{} & \casePhaseOperational{} & \caseAcquisitionStandardTelemetry{} \\
\hline
\textbf{Throughput} & \caseScopeNode{} & \casePhaseOperational{} & \caseAcquisitionCustomInstrumentation{}* \\
\hline
\textbf{Speedup} & \caseScopeMulti{} & \casePhaseExperimental{} & \caseAcquisitionStandardTelemetry{} \\
\hline
\textbf{Parallel Efficiency} & \caseScopeMulti{} & \casePhaseExperimental{} & \caseAcquisitionStandardTelemetry{} \\
\hline
\textbf{Resource Utilization} & \caseScopeNode{} & \casePhaseOperational{} & \caseAcquisitionStandardTelemetry{} \\
\hline
\textbf{Buffer Utilization} & \caseScopeNode{} & \casePhaseOperational{} & \caseAcquisitionStandardTelemetry{} \\
\hline
\textbf{Availability} & \caseScopeNode{} & \casePhaseOperational{} & \caseAcquisitionStandardTelemetry{} \\
\hline
\textbf{Reliability} & \caseScopeNode{} & \casePhaseOperational{} & \caseAcquisitionStandardTelemetry{} \\
\hline
\textbf{Scalability} & \caseScopeMulti{} & \casePhaseExperimental{} & \caseAcquisitionCustomInstrumentation{}* \\
\hline
\textbf{Elasticity} & \caseScopeMulti{} & \casePhaseExperimental{} & \caseAcquisitionExperimentalInstrumentation{} \\
\hline
\textbf{Energy Consumption} & \caseScopeNode{} & \casePhaseOperational{} & \caseAcquisitionExperimentalInstrumentation{} \\
\hline
\textbf{Energy Efficiency} & \caseScopeNode{} & \casePhaseExperimental{} & \caseAcquisitionExperimentalInstrumentation{} \\
\hline
\textbf{Concurrency} & \caseScopeNode{} & \casePhaseOperational{} & \caseAcquisitionStandardTelemetry{} \\
\hline
\textbf{Load Balancing} & \caseScopeMulti{} & \casePhaseOperational{} & \caseAcquisitionStandardTelemetry{} \\
\hline
\textbf{Task Completion Ratio} & \caseScopeNode{} & \casePhaseOperational{} & \caseAcquisitionCustomInstrumentation{}* \\
\hline
\textbf{Cost Efficiency} & \caseScopeNode{} & \casePhaseExperimental{}* & \caseAcquisitionExperimentalInstrumentation{} \\
\hline
\textbf{Comp. Cont. Efficiency} & \caseScopeMulti{} & \casePhaseExperimental{} & \caseAcquisitionExperimentalInstrumentation{} \\
\hline
\end{tabular}
\end{table}

Let $\mathcal{N}$ denote the set of computing-capable nodes in $\mathcal{S}$, such that
\begin{equation}
    \mathcal{N} = \mathcal{C} \cup \mathcal{F} \cup \mathcal{E} \cup \mathcal{M} \cup \Gamma,
\end{equation}
where $|\mathcal{N}|=\zeta$. Let $\mathcal{T}=\{T_1,T_2,\ldots,T_q\}$ denote the set of computational tasks generated by the data producers in $\mathcal{S}$ during an observation window $\Delta t=[t_0,t_1]$. Each task $T_x \in \mathcal{T}$ can be represented as
\begin{equation}\label{eqTx}
    T_x = \langle \omega_x, d_x, \tau_x, \delta_x \rangle,
\end{equation}
where $\omega_x$ denotes the computational workload, $d_x$ denotes the input data size, $\tau_x$ denotes the task arrival time, and $\delta_x$ denotes the task deadline. A task $T_x$ may be executed on one or more computing nodes depending on the scheduling, offloading, and resource allocation mechanism. Let $x_{ij}$ be a binary assignment variable, where
\begin{equation}
    x_{ij} =
    \begin{cases}
    1, & \text{if task } T_i \text{ is assigned to computing node } j,\\
    0, & \text{otherwise}.
    \end{cases}
\end{equation}
The following subsections define computing-level performance metrics suitable for DCCS.

\subsection{Execution and Task Completion Time}

Execution time is one of the most fundamental computing-level performance metrics. Let $T_i$ be a task assigned to computing node $j \in \mathcal{N}$. The execution time can be expressed as
\begin{equation}
    ET_{ij} = \frac{\omega_i}{\mu_j},
\end{equation}
where $\omega_i$ is the computational workload of task $T_i$ and $\mu_j$ is the processing capacity of node $j$. In heterogeneous DCCS, $\mu_j$ varies significantly across IoT, edge, fog, mobile, and cloud nodes.

\textbf{Task Completion Time (TCT)}  measures the total time required to complete a task, including queueing, communication, execution, and result delivery delays. The total response time of task $T_i$ can be represented as
\begin{equation}
    TCT_i = QT_i + CT_i + ET_i + OT_i,
\end{equation}
where $QT_i$ is the queueing time, $CT_i$ is the communication time, $ET_i$ is the execution time, and $OT_i$ is the output transmission or result-return time. If task $T_i$ is executed on node $j$, then
\begin{equation}
    TCT_{i} = QT_{i} + \frac{d_i}{B_{sj}} + \frac{\omega_i}{\mu_j} + \frac{r_i}{B_{jd}},
\end{equation}
where $B_{sj}$ is the bandwidth between the source node and node $j$, $r_i$ is the result size, and $B_{jd}$ is the bandwidth between node $j$ and the destination node.

The average response time of all completed tasks in $\Delta t$ is given by
\begin{equation}
    \overline{TCT} = \frac{1}{|\mathcal{T}^{c}|}\sum_{T_i \in \mathcal{T}^{c}} RT_i,
\end{equation}
where $\mathcal{T}^{c}$ is the set of completed tasks.

As summarized in Table ~\ref{tab:caseComputationMetrics}, the \textbf{Execution Time} metric measures the time required to complete a task.
% Acquisition scope
The minimal acquisition \textbf{scope} is \caseScopeNode{} because a task execution can be completed within a single computational node.
% Acquisition phase
The acquisition occurs during the \casePhaseOperational{} \textbf{phase} in order to detect execution delays and performance degradation during task execution.
% Acquisition method
The acquisition \textbf{method} is \caseAcquisitionStandardTelemetry{} because trace and span durations are already exposed by standard instrumentation frameworks such as OpenTelemetry.
% \vspace{0.4em}% Definition (short)
% Acquisition scope
Similarly, in $TCT$, the minimal acquisition \textbf{scope} is \caseScopeNode{} because the total elapsed time between task submission and completion can be measured locally by the requesting node, regardless of where the task is executed.
% Acquisition phase
The acquisition occurs during the \casePhaseOperational{} \textbf{phase} in order to detect performance degradation not only at the node level but also along the communication path between services.
% Acquisition method
The acquisition \textbf{method} is \caseAcquisitionStandardTelemetry{} because full request durations can be obtained from traces exposed by standard instrumentation frameworks such as OpenTelemetry.

\subsection{Throughput}
% The \textbf{Throughput} metric measures the number of tasks successfully completed per unit of time.
Throughput~\cite{CHAINTREAU200257} measures the number of tasks successfully completed per unit time. In DCCS, throughput can be measured at the node level, layer level, or system level. The system-level throughput during $\Delta t$ is defined as
\begin{equation}
    TH_{\mathcal{S}} = \frac{|\mathcal{T}^{c}|}{\Delta t},
\end{equation}
where $|\mathcal{T}^{c}|$ is the number of completed tasks.

The node-level throughput of node $j$ is
\begin{equation}
    TH_j = \frac{|\mathcal{T}^{c}_j|}{\Delta t},
\end{equation}
where $\mathcal{T}^{c}_j$ is the set of tasks completed by node $j$.

Layer-wise throughput can be computed as
\begin{equation}
    TH_{\mathcal{L}} = \frac{1}{\Delta t}\sum_{j \in \mathcal{L}} |\mathcal{T}^{c}_j|,
\end{equation}
where $\mathcal{L} \in \{\mathcal{C},\mathcal{F},\mathcal{E},\mathcal{M},\Gamma\}$. This formulation helps identify which layer contributes most to task execution. Further, the minimal acquisition \textbf{scope} is \caseScopeNode{} because it can be obtained at the individual node level, although it can also apply to multiple nodes or layers.
% Acquisition phase
The acquisition occurs during the \casePhaseOperational{} \textbf{phase} in order to detect performance degradation.
% Acquisition method
The acquisition \textbf{method} is \caseAcquisitionCustomInstrumentation{} because the application must define what constitutes a successfully completed task (e.g., within a given execution time or meeting a specific quality criterion). However, if task completion is represented by a simple event already exposed through standard metrics, \caseAcquisitionStandardTelemetry{} is sufficient.

\subsection{Speedup}

Speedup measures how much faster a task or workload is executed using a distributed or parallel computing configuration compared with a baseline configuration \cite{aczel1997new}. For a workload $\mathcal{W}$, speedup is defined as
\begin{equation}
    S_p = \frac{ET_1}{ET_p},
\end{equation}
where $ET_1$ is the execution time using one baseline computing node and $ET_p$ is the execution time using $p$ computing nodes.

For DCCS, if a task set $\mathcal{T}$ is executed using $\zeta$ computing nodes, the system-level speedup can be expressed as
\begin{equation}
    S_{\zeta} = \frac{T_{\text{single}}}{T_{\text{dccs}}},
\end{equation}
where $T_{\text{single}}$ is the execution time using a single reference node, and $T_{\text{dccs}}$ is the execution time using the available computing continuum.

Amdahl's law can be used to estimate the theoretical speedup when a fraction $\alpha$ of the workload can be parallelized \cite{amdahl1967validity}. It is given by
\begin{equation}\label{eqAmdhals}
    S_{\text{Amdahl}}(\zeta) = \frac{1}{(1-\alpha)+\frac{\alpha}{\zeta}},
\end{equation}
where $0 \leq \alpha \leq 1$. This formulation is suitable when the problem size is fixed.

For scalable workloads, Gustafson's law can be used \cite{gustafson1988reevaluating}. It is expressed as
\begin{equation}
    S_{\text{Gustafson}}(\zeta) = (1-\alpha) + \alpha \zeta.
\end{equation}
This formulation is more suitable when the workload size increases with the available computing resources.

In DCCS, due to communication delay and synchronization overhead, practical speedup is usually lower than theoretical speedup. Therefore, an overhead-aware speedup can be defined as
\begin{equation}
    S_{\text{DCCS}} = 
    \frac{T_{\text{single}}}
    {T_{\text{compute}} + T_{\text{comm}} + T_{\text{sync}} + T_{\text{migration}}},
\end{equation}
where $T_{\text{compute}}$ is the actual computation time, $T_{\text{comm}}$ is communication time, $T_{\text{sync}}$ is synchronization time, and $T_{\text{migration}}$ is task or data migration time. From Table~\ref{tab:caseComputationMetrics}, the minimal acquisition \textbf{scope} is \caseScopeMulti{} because it compares performance data across different deployment configurations involving multiple computational nodes.
% Acquisition phase
The acquisition occurs during the \casePhaseExperimental{} \textbf{phase} because the metric is primarily used to compare deployment configurations in order to identify the most efficient system design or resource allocation. Although it could be evaluated during operation if a baseline is preconfigured, continuous runtime comparison between configurations is generally impractical.
% Acquisition method
The acquisition \textbf{method} is \caseAcquisitionStandardTelemetry{} because the metric is calculated from execution times, which, as previously explained, can be readily obtained through standard instrumentation.

\subsection{Parallel Efficiency}
The \textbf{Parallel Efficiency} metric measures how close the system's speedup is to the ideal theoretical linear acceleration expected from the number of allocated computing nodes. 
% Parallel efficiency measures how effectively the available computing nodes contribute to speedup. 
It is defined as
\begin{equation}
    PE_{\zeta} = \frac{S_{\zeta}}{\zeta}.
\end{equation}
The value of $PE_{\zeta}$ lies in the range $(0,1]$ for most practical systems. A value close to $1$ indicates near-ideal resource usage, while a lower value indicates communication overhead, synchronization delay, load imbalance, or resource underutilization.

For heterogeneous DCCS, a weighted parallel efficiency can be defined as
\begin{equation}
    WPE = \frac{S_{\zeta}}{\sum_{j \in \mathcal{N}} w_j},
\end{equation}
where $w_j$ represents the normalized computing capacity of node $j$. This is more suitable than simple node counting when the system contains heterogeneous devices such as IoT, edge, fog, mobile, and cloud nodes. Further, the acquisition minimal \textbf{scope} is \caseScopeMulti{} because $PE_{\zeta}$ requires performance data across multiple distributed resources and configurations to calculate the resulting resource usage efficiency.
% Acquisition phase
The acquisition occurs during the \casePhaseExperimental{} \textbf{phase} because it depends on the speedup obtained from comparing different deployment configurations and is therefore primarily intended for design-time performance analysis rather than continuous operational monitoring.
% Acquisition method
The acquisition \textbf{method} is \caseAcquisitionStandardTelemetry{} because it is derived from the speedup and the number of allocated computing nodes, both of which are obtained from data already available through standard telemetry.

\subsection{Resource Utilization}

Resource utilization measures how much of the available computing resources (such as CPU, memory, storage, etc.,) are actually used during a given observation window (e.g., \cite{alexander2009system,murturi2026performance}). The CPU utilization of node $j$ can be defined as
\begin{equation}
    U^{cpu}_{j} = \frac{T^{busy}_{j}}{\Delta t},
\end{equation}
where $T^{busy}_{j}$ is the total time during which the CPU of node $j$ is busy.

The average CPU utilization of the whole DCCS is
\begin{equation}
    U^{cpu}_{\mathcal{S}} = \frac{1}{\zeta}\sum_{j \in \mathcal{N}} U^{cpu}_{j}.
\end{equation}

If the computing nodes have different capacities, weighted CPU utilization can be defined as
\begin{equation}
    WU^{cpu}_{\mathcal{S}} =
    \frac{\sum_{j \in \mathcal{N}} \mu_j U^{cpu}_{j}}
    {\sum_{j \in \mathcal{N}} \mu_j}.
\end{equation}

Memory utilization of node $j$ is defined as
\begin{equation}
    U^{mem}_{j} = \frac{M^{used}_{j}}{M^{total}_{j}},
\end{equation}
where $M^{used}_{j}$ is the used memory and $M^{total}_{j}$ is the total available memory.

Storage utilization is defined as
\begin{equation}
    U^{sto}_{j} = \frac{S^{used}_{j}}{S^{total}_{j}}.
\end{equation}

A composite resource utilization metric can be defined as
\begin{equation}
    U^{res}_{j} =
    ya_1 U^{cpu}_{j}
    + ya_2 U^{mem}_{j}
    + ya_3 U^{sto}_{j}
    + ya_4 U^{net}_{j},
\end{equation}
where $U^{net}_{j}$ is network utilization and $ya_1+ya_2+ya_3+ya_4=1$. The weights can be adjusted according to the application requirements. The acquisition minimal \textbf{scope} is \caseScopeNode{} because $U^{res}_{j}$ metrics are isolated and measured directly within each independent computing node.
% Acquisition phase
The acquisition occurs during the \casePhaseOperational{} \textbf{phase} in order to monitor the operational state of the nodes and detect potential resource exhaustion in real time.
% Acquisition method
The acquisition \textbf{method} is \caseAcquisitionStandardTelemetry{} because infrastructure indicators like CPU load and memory usage are natively exposed by the operating systems (OS).

\subsection{Buffer Utilization}

Buffer utilization is important for DCCS because data may be temporarily stored at IoT, edge, fog, or cloud nodes before processing. Let $B^{used}_{j}(t)$ be the used buffer size of node $j$ at time $t$, and $B^{max}_{j}$ be the maximum buffer capacity \cite{shipman2007investigations}. The instantaneous buffer utilization is
\begin{equation}
    U^{buf}_{j}(t) = \frac{B^{used}_{j}(t)}{B^{max}_{j}}.
\end{equation}

The average buffer utilization during $\Delta t$ is
\begin{equation}
    \overline{U}^{buf}_{j} =
    \frac{1}{\Delta t}
    \int_{t_0}^{t_1}
    \frac{B^{used}_{j}(t)}{B^{max}_{j}} dt.
\end{equation}

In discrete monitoring intervals, this can be written as
\begin{equation}
    \overline{U}^{buf}_{j} =
    \frac{1}{H}
    \sum_{h=1}^{H}
    \frac{B^{used}_{j}(h)}{B^{max}_{j}},
\end{equation}
where $H$ is the number of monitoring samples.

A high buffer utilization may indicate efficient buffer usage, but very high values can lead to congestion and packet/task dropping. Therefore, buffer overflow probability can be defined as
\begin{equation}
    P^{overflow}_{j} =
    \frac{N^{drop}_{j}}{N^{arr}_{j}},
\end{equation}
where $N^{drop}_{j}$ is the number of dropped tasks or packets and $N^{arr}_{j}$ is the total number of arrivals at node $j$. The $U^{buf}_{j}(t)$ metric measures the proportion of the available buffer capacity currently in use. Since a buffer can be considered a computational resource managed by a node, its acquisition requirements are equivalent to those of the $U^{res}_{j}$ metric: the minimal acquisition \textbf{scope} is \caseScopeNode{}, the acquisition occurs during the \casePhaseOperational{} \textbf{phase}, and the acquisition \textbf{method} is \caseAcquisitionStandardTelemetry{}.

\subsection{Availability}

Availability measures the probability that a computing node or the whole DCCS is operational at a given time. For node $j$, availability can be expressed using mean time to failure (MTTF) and mean time to repair (MTTR) as
\begin{equation}
    A_j = \frac{MTTF_j}{MTTF_j + MTTR_j}.
\end{equation}

If the failure and repair rates are represented as $\lambda_j$ and $\mu_j^{r}$, respectively, then availability can also be written as
\begin{equation}
    A_j = \frac{\mu_j^{r}}{\lambda_j + \mu_j^{r}}.
\end{equation}

The average availability of the DCCS is
\begin{equation}
    A_{\mathcal{S}} =
    \frac{1}{\zeta}
    \sum_{j \in \mathcal{N}} A_j.
\end{equation}

For capacity-aware availability, the formulation can be extended as
\begin{equation}
    WA_{\mathcal{S}} =
    \frac{\sum_{j \in \mathcal{N}} \mu_j A_j}
    {\sum_{j \in \mathcal{N}} \mu_j}.
\end{equation}

If the DCCS is considered operational only when at least one cloud, one fog/edge, and one data-producing layer are available, system availability can be represented as
\begin{equation}
    A_{\mathcal{S}} =
    A_{\mathcal{C}}
    \cdot
    \left[1-(1-A_{\mathcal{F}})(1-A_{\mathcal{E}})\right]
    \cdot
    A_{\rho},
\end{equation}
where $A_{\mathcal{C}}$, $A_{\mathcal{F}}$, $A_{\mathcal{E}}$, and $A_{\rho}$ denote the availability of the cloud, fog, edge, and data-producing components, respectively. The acquisition minimal \textbf{scope} is \caseScopeNode{} because availability can be estimated independently for each computing node.
% Acquisition phase
The acquisition occurs during the \casePhaseOperational{} \textbf{phase} because $A_j$ is useful for detecting nodes that experience frequent outages in production. Note that meaningful availability estimates require sufficient historical operational data to produce reliable results.
% Acquisition method
The acquisition \textbf{method} is \caseAcquisitionStandardTelemetry{} because operational status information can be collected through standard monitoring mechanisms, such as heartbeat signals, which provide the data required for availability estimation.

\subsection{Reliability}

Reliability measures the probability that a computing node or system operates without failure for a given time duration. If node $j$ has a constant failure rate $\lambda_j$, its reliability over time $t$ is
\begin{equation}
    \mathbb{R}_j(t) = e^{-\lambda_j t}.
\end{equation}

The reliability of a task execution path $P_i$ can be defined as
\begin{equation}
    \mathbb{R}(P_i) = \prod_{j \in P_i} \mathbb{R}_j(t),
\end{equation}
where $P_i$ includes the computing and communication nodes involved in executing task $T_i$.

For replicated task execution across multiple nodes, reliability can be improved as
\begin{equation}
  \mathbb{R}^{rep}_{i} =
    1 -
    \prod_{j \in \mathcal{R}_i}
    \left(1-\mathbb{R}_j(t)\right),
\end{equation}
where $\mathcal{R}_i$ is the set of nodes selected for replicated execution of task $T_i$. The acquisition minimal \textbf{scope} is \caseScopeNode{} because reliability is estimated independently for each computing node based on its failure history.
% Acquisition phase
As shown in Table~\ref{tab:caseComputationMetrics}, the acquisition occurs during the \casePhaseOperational{} \textbf{phase} because reliability helps identify nodes that are prone to failures during production. Note that, besides requiring historical failure data, the metric must be evaluated over a predefined time interval. Therefore, real-time reliability monitoring is only practical when this temporal parameter is defined in advance.
% Acquisition method
The acquisition \textbf{method} is \caseAcquisitionStandardTelemetry{} because failure events and operational time can be collected through standard monitoring mechanisms, such as heartbeat signals, and later used to compute reliability estimates.

\subsection{Scalability}

Scalability measures the ability of the DCCS to maintain acceptable performance when workload or system size increases. It can be measured through speedup, scaleup, sizeup, and resource expansion efficiency. This performance metric is generally used in several computing architectures in the literature including, cloud, or IoT. However, this is very relevant in the context of DCCS as well. 

The throughput-based scalability from configuration $a$ to configuration $b$ can be defined as
\begin{equation}
    SC_{TH} =
    \frac{TH_b}{TH_a}
    \cdot
    \frac{\mathscr{R}_a}{\mathscr{R}_b},
\end{equation}
where $TH_a$ and $TH_b$ are throughputs under configurations $a$ and $b$, and $\mathscr{R}_a$ and $\mathscr{R}_b$ are the amount of provisioned resources.

A scalability value close to $1$ indicates near-linear scalability. A value lower than $1$ indicates that increasing resources does not proportionally improve throughput.

The workload scalability can be represented as
\begin{equation}
    SC_{W} =
    \frac{W_b/W_a}{RT_b/RT_a},
\end{equation}
where $W_a$ and $W_b$ are workload intensities, while $RT_a$ and $RT_b$ are the corresponding response times. If the workload increases but response time remains stable, the system is considered scalable.

A deadline-aware scalability metric can be defined as
\begin{equation}
    SC_{\delta} =
    \frac{|\mathcal{T}^{c}_{\delta}|}{|\mathcal{T}|},
\end{equation}
where $\mathcal{T}^{c}_{\delta}$ is the set of tasks completed within their deadlines. This is useful for real-time DCCS applications. The minimal acquisition \textbf{scope} is \caseScopeMulti{} because scalability is evaluated by comparing throughput across multiple deployment configurations.
% Acquisition phase
The acquisition occurs during the \casePhaseExperimental{} \textbf{phase} because scalability is primarily assessed through controlled experiments that compare different configurations in order to identify the most effective system design.
% Acquisition method
The acquisition \textbf{method} is \caseAcquisitionCustomInstrumentation{} because, although resource metrics are available through \caseAcquisitionStandardTelemetry{}, throughput may require custom instrumentation, as previously explained.

\subsection{Elasticity}
The \textbf{Elasticity} metric measures the system's capacity to adapt its resource provisioning dynamically to match varying workload demands over time \cite{beltran2016defining,7937885}. Unlike scalability, elasticity includes the temporal behavior of resource provisioning and de-provisioning \cite{herbst2013elasticity,beltran2016defining, murturi2022decent,sedlak2024equilibrium}. This performance metric extended according to the recent advancements in computing paradigms under novel performance metrics in section \ref{subsec:Equilibrium}. 

Let $D(t)$ denote the resource demand at time $t$, and $P(t)$ denote the provisioned resources. Resource provisioning accuracy can be measured as
\begin{equation}
    EA(t) =
    1 -
    \frac{|P(t)-D(t)|}{D(t)}.
\end{equation}

The average elasticity accuracy over $\Delta t$ is
\begin{equation}
    \overline{EA} =
    1 -
    \frac{1}{H}
    \sum_{h=1}^{H}
    \frac{|P(h)-D(h)|}{D(h)}.
\end{equation}

Over-provisioning and under-provisioning can be measured separately as
\begin{equation}
    OP =
    \frac{1}{H}
    \sum_{h=1}^{H}
    \max\left(0,\frac{P(h)-D(h)}{D(h)}\right),
\end{equation}
and
\begin{equation}
    UP =
    \frac{1}{H}
    \sum_{h=1}^{H}
    \max\left(0,\frac{D(h)-P(h)}{D(h)}\right).
\end{equation}

The elasticity reaction time is defined as
\begin{equation}
    ERT = t_{\text{stable}} - t_{\text{change}},
\end{equation}
where $t_{\text{change}}$ is the time when workload changes and $t_{\text{stable}}$ is the time when the system reaches a stable resource configuration.

A combined elasticity score can be defined as
\begin{equation}
    ES =
    \eta_1 \overline{EA}
    +
    \eta_2 (1-OP)
    +
    \eta_3 (1-UP)
    +
    \eta_4 \frac{1}{1+ERT},
\end{equation}
where $\eta_1+\eta_2+\eta_3+\eta_4=1$. 
% Acquisition scope
The acquisition minimal \textbf{scope} is \caseScopeMulti{} because it requires evaluating changes in resource configurations across multiple distributed nodes.
% Acquisition phase
The acquisition \textbf{phase} is \casePhaseExperimental{} because analyzing the speed and precision of provisioning adaptation requires controlled stress or workload scaling experiments.
% Acquisition method
The acquisition \textbf{method} is \caseAcquisitionExperimentalInstrumentation{} because, although provisioned resources are readily available through standard telemetry, estimating the actual resource demand often requires experimental models or external evaluation procedures that are not part of conventional monitoring.

\subsection{Energy Consumption and Efficiency}

Energy consumption and energy efficiency \cite{muralidhar2022energy,rong2016optimizing,prieto2025evolution,alsharif2025survey} are critical performance metrics in several computing paradigms, and  DCCS is not exceptional because many IoT, sensor, edge, and mobile devices are battery-powered \cite{govori2023comprehensive} or energy dependent. It is noteworthy that, several works in the literature oscillate among measuring energy consumption or energy efficiency \cite{8882255}. The energy consumed by node $j$ during $\Delta t$ can be computed as
\begin{equation}
    E_j =
    \int_{t_0}^{t_1} P_j(t) dt,
\end{equation}
where $P_j(t)$ is the power consumption of node $j$ at time $t$.

In discrete form,
\begin{equation}
    E_j =
    \sum_{h=1}^{H}
    P_j(h)\Delta h.
\end{equation}

The total energy consumption of the DCCS is
\begin{equation}
    E_{\mathcal{S}} =
    \sum_{j \in \mathcal{N}} E_j.
\end{equation}

The energy consumed for task $T_i$ executed on node $j$ can be approximated as
\begin{equation}
    E_{ij} =
    P^{cpu}_{j} ET_{ij}
    +
    P^{comm}_{j} CT_{ij}
    +
    P^{idle}_{j} IT_{ij},
\end{equation}
where $P^{cpu}_{j}$, $P^{comm}_{j}$, and $P^{idle}_{j}$ denote computation, communication, and idle power consumption, respectively. The minimal acquisition \textbf{scope} is \caseScopeNode{} because energy consumption can be measured independently for each computing node.
% Acquisition phase
The acquisition occurs during the \casePhaseOperational{} \textbf{phase} because monitoring energy consumption is critical for battery-powered DCCS nodes, where energy availability directly limits operating time.
% Acquisition method
The acquisition \textbf{method} is \caseAcquisitionExperimentalInstrumentation{} because accurate energy measurements typically require hardware energy sensors or platform-specific monitoring interfaces. When direct measurements are unavailable, such as in cloud environments, a rough estimation can be obtained from resource utilization and the processor's TDP, as proposed in \cite{OTERO2025101100}.

% \subsection{Energy Efficiency}

\textbf{Energy efficiency} measures the amount of useful work completed per unit energy. It can be defined as
\begin{equation}
    EE_{\mathcal{S}} =
    \frac{|\mathcal{T}^{c}|}{E_{\mathcal{S}}},
\end{equation}
where $|\mathcal{T}^{c}|$ is the number of completed tasks.

If workload size is considered, energy efficiency can be expressed as
\begin{equation}
    EE^{work}_{\mathcal{S}} =
    \frac{\sum_{T_i \in \mathcal{T}^{c}} \omega_i}
    {E_{\mathcal{S}}}.
\end{equation}

For deadline-sensitive applications, useful energy efficiency can be defined as
\begin{equation}
    EE^{\delta}_{\mathcal{S}} =
    \frac{|\mathcal{T}^{c}_{\delta}|}
    {E_{\mathcal{S}}},
\end{equation}
where $\mathcal{T}^{c}_{\delta}$ denotes the set of tasks completed within their deadlines.

Energy-delay product (EDP) can also be used to jointly evaluate energy and delay:
\begin{equation}
    EDP_i = E_i \times RT_i.
\end{equation}
A lower EDP indicates better energy-delay performance. 
% \vspace{0.4em}% Definition (short)
% The \textbf{Energy Efficiency} metric measures the amount of useful work completed per unit of energy consumed.
% Acquisition scope
The minimal acquisition \textbf{scope} is \caseScopeNode{} because the completed workload and the corresponding energy consumption can be measured for an individual computing node.
% Acquisition phase
The acquisition occurs during the \casePhaseExperimental{} \textbf{phase} because energy efficiency is primarily used to evaluate and compare system designs under controlled conditions rather than to support operational decisions, although it can also be computed during operation when energy consumption measurements are available.
% Acquisition method
The acquisition \textbf{method} is \caseAcquisitionExperimentalInstrumentation{} because it inherits the acquisition requirements of the Energy Consumption metric, previously discussed.

\subsection{Concurrency}

Concurrency measures the ability of the computer systems to handle multiple tasks simultaneously~\cite{Lamport1983con1,Lamport1983con2,Lamport1983con3}. In DCCS, let $\mathbb{Q}(t)$ be the number of tasks being processed or waiting in the system at time $t$. The average concurrency level is
\begin{equation}
    CL =
    \frac{1}{\Delta t}
    \int_{t_0}^{t_1} \mathbb{Q}(t)dt.
\end{equation}

Using Little's law, the average number of tasks in the system can be related to throughput and response time as
\begin{equation}
    \mathbb{Q} = \Lambda W,
\end{equation}
where $\Lambda$ is the task arrival rate and $W$ is the average time a task spends in the system.

For DCCS, this can be written as
\begin{equation}
    CL_{\mathcal{S}} =
    \Lambda_{\mathcal{S}} \overline{RT},
\end{equation}
where $\Lambda_{\mathcal{S}}$ is the system-level task arrival rate and $\overline{RT}$ is the average response time.

The maximum concurrency capacity can be estimated as
\begin{equation}
    CC_{\mathcal{S}} =
    \sum_{j \in \mathcal{N}} q_j,
\end{equation}
where $q_j$ is the maximum number of tasks that node $j$ can execute concurrently.

The concurrency utilization is then
\begin{equation}
    U^{con}_{\mathcal{S}} =
    \frac{CL_{\mathcal{S}}}{CC_{\mathcal{S}}}.
\end{equation}
The minimal acquisition \textbf{scope} is \caseScopeNode{} because the number of concurrent tasks can be measured independently for each computing node and later aggregated to characterize the entire system.
% Acquisition phase
The acquisition occurs during the \casePhaseOperational{} \textbf{phase} because monitoring concurrency helps identify workload saturation during system operation.
% Acquisition method
The acquisition \textbf{method} is \caseAcquisitionStandardTelemetry{} because the number of concurrently executing or queued tasks can typically be obtained from existing runtime or infrastructure telemetry.

\subsection{Load Balancing}

Load balancing measures how evenly tasks or workloads are distributed across computing nodes \cite{10960753}. Let $Load_j$ denote the load assigned to node $j$. This load can represent CPU usage, number of tasks, queue length, or workload size. The average load is
\begin{equation}
    \overline{Load} =
    \frac{1}{\zeta}
    \sum_{j \in \mathcal{N}} Load_j.
\end{equation}

The load imbalance degree can be defined as
\begin{equation}
    LID =
    \frac{\sqrt{\frac{1}{\zeta}\sum_{j \in \mathcal{N}}(Load_j-\overline{Load})^2}}
    {\overline{Load}}.
\end{equation}
A lower $LID$ indicates better load balancing.

Jain's fairness index can also be used to measure load distribution fairness \cite{jain1984fairness}. It is defined as
\begin{equation}
    JFI =
    \frac{\left(\sum_{j \in \mathcal{N}} Load_j\right)^2}
    {\zeta \sum_{j \in \mathcal{N}} Load_j^2}.
\end{equation}
The value of $JFI$ lies in $(0,1]$, where $1$ indicates perfectly balanced load distribution.

For heterogeneous DCCS, load should be normalized by node capacity. Therefore, normalized load can be defined as
\begin{equation}
    NLoad_j =
    \frac{Load_j}{\mu_j}.
\end{equation}
The normalized load imbalance degree is
\begin{equation}
    NLID =
    \frac{\sqrt{\frac{1}{\zeta}\sum_{j \in \mathcal{N}}(NLoad_j-\overline{NLoad})^2}}
    {\overline{NLoad}},
\end{equation}
where $\overline{NLoad}$ is the average normalized load. As shown in Table~\ref{tab:caseComputationMetrics}, The minimal acquisition \textbf{scope} is \caseScopeMulti{} because load balancing is evaluated by comparing the load assigned to multiple computing nodes.
% Acquisition phase
The acquisition occurs during the \casePhaseOperational{} \textbf{phase} because it enables the detection of workload imbalances during system operation.
% Acquisition method
The acquisition \textbf{method} is \caseAcquisitionStandardTelemetry{} because node load indicators, such as resource utilization, queue length, or task count, are typically available through existing instrumentation.

\subsection{Task Completion Ratio}

Task completion ratio measures the fraction of generated tasks that are successfully completed or it measures the proportion of successfully completed tasks with respect to the total number of submitted tasks.
 It is defined as
\begin{equation}
    TCR =
    \frac{|\mathcal{T}^{c}|}{|\mathcal{T}|}.
\end{equation}

For deadline-constrained applications, deadline-aware task completion ratio is more suitable:
\begin{equation}
    DTCR =
    \frac{|\mathcal{T}^{c}_{\delta}|}{|\mathcal{T}|}.
\end{equation}

If failed tasks are represented by $\mathcal{T}^{f}$, then task failure ratio is
\begin{equation}
    TFR =
    \frac{|\mathcal{T}^{f}|}{|\mathcal{T}|}.
\end{equation}
% Acquisition scope
The minimal acquisition \textbf{scope} is \caseScopeNode{} because task submissions and completions can be measured independently for each computing node.
% Acquisition phase
The acquisition occurs during the \casePhaseOperational{} \textbf{phase} because monitoring the task completion ratio helps detect execution failures or service degradation during system operation.
% Acquisition method
The acquisition \textbf{method} is \caseAcquisitionCustomInstrumentation{} because the application must define what constitutes a successfully completed task, as previously discussed for the Throughput metric.

\subsection{Cost Efficiency}
This metric measures the amount of useful work completed with respect to the operational cost of the computing resources.
Although cost is not purely a computing-level metric, it is strongly related to computing performance in DCCS because cloud, fog, edge, and mobile resources have different deployment and operational costs. Let $Cost_j$ denote the operational cost of using node $j$ during $\Delta t$. The total system cost is
\begin{equation}
    Cost_{\mathcal{S}} =
    \sum_{j \in \mathcal{N}} Cost_j.
\end{equation}

The cost per completed task is
\begin{equation}
    CPT =
    \frac{Cost_{\mathcal{S}}}{|\mathcal{T}^{c}|}.
\end{equation}

The cost-performance efficiency can be defined as
\begin{equation}
    CPE =
    \frac{TH_{\mathcal{S}}}{Cost_{\mathcal{S}}}.
\end{equation}

A combined energy-cost efficiency metric can be expressed as
\begin{equation}
    ECE =
    \frac{|\mathcal{T}^{c}_{\delta}|}
    {w_1 E_{\mathcal{S}} + w_2 Cost_{\mathcal{S}}},
\end{equation}
where $w_1$ and $w_2$ are weighting factors. 
% Acquisition scope
The minimal acquisition \textbf{scope} is \caseScopeNode{} because the operational cost can be estimated independently for each computing node and later aggregated at the system level if needed.
% Acquisition phase
The acquisition occurs during the \casePhaseExperimental{}* \textbf{phase} because cost efficiency is primarily used to evaluate deployment strategies. However, it can also be computed during the \casePhaseOperational{} phase if cost estimation is continuously available with acceptable overhead.
% Acquisition method
The acquisition \textbf{method} is \caseAcquisitionExperimentalInstrumentation{} because estimating operational cost requires integrating cost models or pricing information, for which no standard telemetry mechanism currently exists.

\section{Network-level Performance Measurement}\label{sec:NetworkLevel}
Network-level performance measurement evaluates the efficiency, reliability, and quality of communication among heterogeneous devices in a DCCS. In contrast to computing-level performance, which focuses on task execution and resource usage, network-level performance focuses on data transmission among sensor nodes, IoT devices, mobile devices, edge nodes, fog nodes, and cloud/data centers. Since DCCS involves geographically distributed, mobile, and resource-constrained components (i.e., whose intelligent management increasingly relies on AI/ML-driven orchestration across the continuum~\cite{murturi2026intelligent}) network-level metrics are essential for measuring latency, bandwidth usage, packet delivery, throughput, service delay, communication reliability, and network availability. Table \ref{tab:network_metrics} summarizes the acquisition requirements associated with each network-level metric. 
\begin{table}[t]
\centering
\caption{Acquisition requirements for network-level metrics.}
\label{tab:network_metrics}
\begin{tabular}{|l|c|c|c|}

\hline
\textbf{Metric} & \textbf{Scope} & \textbf{Phase} & \textbf{Method} \\ \hline

\hline
\textbf{Network Latency} & \caseScopeMulti{} & \casePhaseOperational{} & \caseAcquisitionStandardTelemetry{} \\
\hline
\textbf{Jitter} & \caseScopeNode{} & \casePhaseOperational{} & \caseAcquisitionStandardTelemetry{} \\
\hline
\textbf{Bandwidth Utilization} & \caseScopeNode{} & \casePhaseOperational{} & \caseAcquisitionStandardTelemetry{} \\
\hline
\textbf{Packet Loss Rate} & \caseScopeMulti{} & \casePhaseOperational{} & \caseAcquisitionStandardTelemetry{} \\
\hline
\textbf{Packet Delivery Rate} & \caseScopeMulti{} & \casePhaseOperational{} & \caseAcquisitionStandardTelemetry{} \\
\hline
\textbf{Network Throughput} & \caseScopeNode{} & \casePhaseOperational{} & \caseAcquisitionStandardTelemetry{} \\
\hline
\textbf{Goodput} & \caseScopeNode{} & \casePhaseOperational{} & \caseAcquisitionCustomInstrumentation{} \\
\hline

\textbf{Network Response Time} & \caseScopeNode{} & \casePhaseOperational{} & \caseAcquisitionStandardTelemetry{} \\
\hline

\textbf{Service Time} & \caseScopeNode{} & \casePhaseOperational{} & \caseAcquisitionStandardTelemetry{} \\
\hline

\textbf{Queueing Delay} & \caseScopeNode{} & \casePhaseOperational{} & \caseAcquisitionStandardTelemetry{} \\
\hline

\textbf{Network Congestion Level} & \caseScopeNode{} & \casePhaseOperational{} & \caseAcquisitionStandardTelemetry{} \\
\hline

\textbf{Network Availability} & \caseScopeNode{} & \casePhaseOperational{} & \caseAcquisitionStandardTelemetry{} \\
\hline

\textbf{Network Reliability} & \caseScopeNode{} & \casePhaseOperational{} & \caseAcquisitionStandardTelemetry{} \\
\hline

\textbf{Handover Delay} & \caseScopeNode{}* & \casePhaseOperational{} & \caseAcquisitionStandardTelemetry{} \\
\hline

\textbf{Comm. Energy Consumpt.} & \caseScopeNode{} & \casePhaseOperational{} & \caseAcquisitionExperimentalInstrumentation{} \\
\hline

\textbf{Comm. Energy Efficiency} & \caseScopeNode{} & \casePhaseExperimental{} & \caseAcquisitionExperimentalInstrumentation{} \\
\hline

\textbf{Connection Density} & \caseScopeSystem{} & \casePhaseOperational{} & \caseAcquisitionStandardTelemetry{} \\
\hline

\textbf{Path Stretch} & \caseScopeSystem{} & \casePhaseOperational{} & \caseAcquisitionStandardTelemetry{} \\
\hline

\textbf{Network Cost} & \caseScopeNode{} & \casePhaseExperimental{} & \caseAcquisitionExperimentalInstrumentation{} \\
\hline

\textbf{Network Service Quality} & \caseScopeSystem{} & \casePhaseExperimental{} & \caseAcquisitionExperimentalInstrumentation{} \\
\hline

\end{tabular}
\end{table}

Let $\mathcal{N}^{all}$ denote the set of all devices in $\mathcal{S}$, including computing and non-computing devices:
\begin{equation}
    \mathcal{N}^{all} = \mathcal{C} \cup \mathcal{F} \cup \mathcal{E} \cup \mathcal{M} \cup \Gamma \cup \Psi,
\end{equation}
where $|\mathcal{N}^{all}|=n$. The connection between any two devices $i,j \in \mathcal{N}^{all}$ is represented using the connection matrix $\mathbb{C}_{ij}$:
\begin{equation}
    \mathbb{C}_{ij} =
    \begin{cases}
    1, & \text{if nodes } i \text{ and } j \text{ can communicate directly},\\
    0, & \text{otherwise}.
    \end{cases}
\end{equation}
Let $\mathcal{L}$ denote the set of communication links in $\mathcal{S}$:
\begin{equation}
    \mathcal{L} = \{(i,j) \mid i,j \in \mathcal{N}^{all}, i \neq j, \mathbb{C}_{ij}=1\}.
\end{equation}
For each link $(i,j)\in \mathcal{L}$, let $B_{ij}$ denote the available bandwidth,  $P^{tx}_{ij}$ denote the number of transmitted packets, $P^{rx}_{ij}$ denote the number of successfully received packets, and $P^{loss}_{ij}$ denote the number of lost packets during an observation window $\Delta t=[t_0,t_1]$.

\subsection{Network Latency}

Network latency measures the time required for data to travel from a source node to a destination node \cite{link2000method}. In DCCS, latency is critical because sensor, IoT, edge, and mobile applications often require timely data delivery. For a packet transmitted from node $i$ to node $j$, one-way latency can be expressed as
\begin{equation}
    L_{ij} = t^{rx}_{j} - t^{tx}_{i},
\end{equation}
where $t^{tx}_{i}$ is the transmission time at source node $i$ and $t^{rx}_{j}$ is the reception time at destination node $j$.

The round-trip time (RTT)\cite{hansson2026predictivertocoapusing} between nodes $i$ and $j$ is defined as
\begin{equation}
    RTT_{ij} = t^{reply}_{i} - t^{request}_{i},
\end{equation}
where $t^{request}_{i}$ is the time when node $i$ sends a request and $t^{reply}_{i}$ is the time when node $i$ receives the corresponding reply.

The average network latency over all active communication links is
\begin{equation}
    \overline{L}_{\mathcal{S}} =
    \frac{1}{|\mathcal{L}|}
    \sum_{(i,j)\in \mathcal{L}} L_{ij}.
\end{equation}

For a multi-hop communication path $P_{sd}$ from source $s$ to destination $d$, the end-to-end latency is
\begin{equation}
    L_{sd}^{e2e} =
    \sum_{(i,j)\in P_{sd}}
    \left(
    L^{prop}_{ij}
    +
    L^{trans}_{ij}
    +
    L^{queue}_{ij}
    +
    L^{proc}_{ij}
    \right),
\end{equation}
where $L^{prop}_{ij}$ is propagation delay, $L^{trans}_{ij}$ is transmission delay, $L^{queue}_{ij}$ is queueing delay, and $L^{proc}_{ij}$ is processing delay at intermediate nodes.

The transmission delay for a packet of size $d_p$ over link $(i,j)$ is
\begin{equation}
    L^{trans}_{ij} = \frac{d_p}{B_{ij}}.
\end{equation}
As summarized in Table~\ref{tab:network_metrics}, the minimal acquisition \textbf{scope} is \caseScopeMulti{} because latency is inherently defined between a source and a destination. Although RTT can be measured from a single endpoint, it still requires participation from the remote node, while one-way latency additionally requires synchronized timestamps at both ends.
% Acquisition phase
The acquisition occurs during the \casePhaseOperational{} \textbf{phase} to detect communication delays during system execution.
% Acquisition method
The acquisition \textbf{method} is \caseAcquisitionStandardTelemetry{} because latency is provided by standard networking and observability mechanisms. 

\subsection{Jitter}
Jitter measures the variation in packet delay over time or the variation in packet delay between consecutive packets. It is especially important for real-time DCCS applications such as autonomous vehicles, remote monitoring, industrial control, augmented reality, and video analytics. If $L_{ij}^{p}$ and $L_{ij}^{p-1}$ are the latencies of two consecutive packets transmitted over link $(i,j)$, packet delay variation can be expressed as
\begin{equation}
    J_{ij}^{p} = |L_{ij}^{p} - L_{ij}^{p-1}|.
\end{equation}

The average jitter over $P$ received packets is
\begin{equation}
    \overline{J}_{ij} =
    \frac{1}{P-1}
    \sum_{p=2}^{P}
    |L_{ij}^{p} - L_{ij}^{p-1}|.
\end{equation}

For the complete DCCS, average jitter can be computed as
\begin{equation}
    \overline{J}_{\mathcal{S}} =
    \frac{1}{|\mathcal{L}|}
    \sum_{(i,j)\in \mathcal{L}}
    \overline{J}_{ij}.
\end{equation}

A lower jitter value indicates more stable communication. This metric is commonly used in IP performance measurement, especially for delay-sensitive networked systems \cite{rfc3393}. The minimal acquisition \textbf{scope} is \caseScopeNode{} because packet delay variation can be computed locally from consecutive delay measurements collected by a single endpoint.
% Acquisition phase
The acquisition occurs during the \casePhaseOperational{} \textbf{phase} in order to detect communication instability during system execution.
% Acquisition method
The acquisition \textbf{method} is \caseAcquisitionStandardTelemetry{} because packet delay variation is routinely exposed by standard networking measurement and monitoring tools.

\subsection{Bandwidth Utilization}

Bandwidth utilization measures the fraction of available link capacity that is actually used for data transmission \cite{prasad2003bandwidth}. For a link $(i,j)$, bandwidth utilization is defined as
\begin{equation}
    U^{bw}_{ij} =
    \frac{R_{ij}}{B_{ij}},
\end{equation}
where $R_{ij}$ is the actual data transmission rate and $B_{ij}$ is the maximum available bandwidth of link $(i,j)$.

The average bandwidth utilization of the DCCS is
\begin{equation}
    U^{bw}_{\mathcal{S}} =
    \frac{1}{|\mathcal{L}|}
    \sum_{(i,j)\in \mathcal{L}}
    U^{bw}_{ij}.
\end{equation}

For heterogeneous links, a capacity-weighted bandwidth utilization can be defined as
\begin{equation}
    WU^{bw}_{\mathcal{S}} =
    \frac{
    \sum_{(i,j)\in \mathcal{L}} R_{ij}
    }
    {
    \sum_{(i,j)\in \mathcal{L}} B_{ij}
    }.
\end{equation}

A very low value of $U^{bw}_{ij}$ may indicate underutilized communication resources, while a value close to $1$ may indicate congestion risk. The minimal acquisition \textbf{scope} is \caseScopeNode{} because the transmitted traffic and link capacity can be measured locally at a network interface.
% Acquisition phase
The acquisition occurs during the \casePhaseOperational{} \textbf{phase} in order to detect network resource saturation during system execution.
% Acquisition method
The acquisition \textbf{method} is \caseAcquisitionStandardTelemetry{} because  bandwidth utilization is usually exposed by the OS.

\subsection{Packet Loss Rate}

Packet loss rate measures the fraction of transmitted packets that fail to reach the destination \cite{4456413}. It is an important metric for evaluating communication reliability in DCCS, especially when data are transmitted from low-power sensor and IoT devices over unreliable wireless links.

For a link $(i,j)$, packet loss rate is defined as
\begin{equation}
    PLR_{ij} =
    \frac{P^{loss}_{ij}}{P^{tx}_{ij}},
\end{equation}
where $P^{loss}_{ij}$ is the number of lost packets and $P^{tx}_{ij}$ is the number of transmitted packets.

Since
\begin{equation}
    P^{loss}_{ij} = P^{tx}_{ij} - P^{rx}_{ij},
\end{equation}
packet loss rate can also be written as
\begin{equation}
    PLR_{ij} =
    \frac{P^{tx}_{ij} - P^{rx}_{ij}}{P^{tx}_{ij}}.
\end{equation}

The system-level packet loss rate is
\begin{equation}
    PLR_{\mathcal{S}} =
    \frac{
    \sum_{(i,j)\in \mathcal{L}} 
    \left(P^{tx}_{ij} - P^{rx}_{ij}\right)
    }
    {
    \sum_{(i,j)\in \mathcal{L}} P^{tx}_{ij}
    }.
\end{equation}

A high packet loss rate may increase retransmissions, energy consumption, latency, and task failure probability. The minimal acquisition \textbf{scope} is \caseScopeMulti{} because packet loss is determined by comparing the number of transmitted packets at the sender with the number of received packets at the destination.
% Acquisition phase
The acquisition occurs during the \casePhaseOperational{} \textbf{phase} in order to detect communication reliability degradation during system execution.
% Acquisition method
The acquisition \textbf{method} is \caseAcquisitionStandardTelemetry{} because transmitted and received packet counters are routinely exposed by network interfaces and monitoring frameworks.

\subsection{Packet Delivery Rate}

Packet delivery rate/ratio is complementary to \textbf{Packet Loss Rate} ($PDR = 1 - PLR$) and therefore shares the same acquisition scope, phase, and method. It measures the fraction of transmitted packets that are successfully received at the destination. For a link $(i,j)$, it is defined as
\begin{equation}
    PDR_{ij} =
    \frac{P^{rx}_{ij}}{P^{tx}_{ij}}.
\end{equation}

Since packet delivery and packet loss are complementary, the relationship between packet delivery rate and packet loss rate is
\begin{equation}
    PDR_{ij} = 1 - PLR_{ij}.
\end{equation}

The system-level packet delivery rate is
\begin{equation}
    PDR_{\mathcal{S}} =
    \frac{
    \sum_{(i,j)\in \mathcal{L}} P^{rx}_{ij}
    }
    {
    \sum_{(i,j)\in \mathcal{L}} P^{tx}_{ij}
    }.
\end{equation}

For multi-hop communication, if packets travel through path $P_{sd}$, the end-to-end packet delivery rate can be approximated as
\begin{equation}
    PDR_{sd}^{e2e} =
    \prod_{(i,j)\in P_{sd}} PDR_{ij}.
\end{equation}

\subsection{Network Throughput}

Network throughput measures the amount of useful data successfully delivered per unit time. Unlike bandwidth, which represents the theoretical or provisioned capacity, throughput represents the actual achieved data transmission rate \cite{persico2015measuring,xie2026risk}. For a link $(i,j)$, throughput is defined as
\begin{equation}
    TH^{net}_{ij} =
    \frac{D^{rx}_{ij}}{\Delta t},
\end{equation}
where $D^{rx}_{ij}$ is the amount of successfully received data over link $(i,j)$ during $\Delta t$.

The total network throughput of the DCCS is
\begin{equation}
    TH^{net}_{\mathcal{S}} =
    \frac{
    \sum_{(i,j)\in \mathcal{L}} D^{rx}_{ij}
    }
    {\Delta t}.
\end{equation}

Layer-wise network throughput can be defined as
\begin{equation}
    TH^{net}_{\mathcal{X}} =
    \frac{
    \sum_{i\in \mathcal{X}}
    \sum_{j\in \mathcal{N}^{all}}
    D^{rx}_{ij}
    }
    {\Delta t},
\end{equation}
where $\mathcal{X} \in \{\Psi,\Gamma,\mathcal{M},\mathcal{E},\mathcal{F},\mathcal{C}\}$. This metric helps identify which layer generates or carries the highest amount of traffic. The minimal acquisition \textbf{scope} is \caseScopeNode{} because the amount of successfully received data can be measured locally at the receiving endpoint.
% Acquisition phase
The acquisition occurs during the \casePhaseOperational{} \textbf{phase} in order to detect communication performance during system execution.
% Acquisition method
The acquisition \textbf{method} is \caseAcquisitionStandardTelemetry{} because received byte counters and data rates are routinely exposed by network interfaces and OS.

\subsection{Goodput}

Goodput measures the amount of useful application-level data delivered per unit time \cite{basney1999improving,wongpanich2025machine}. Unlike throughput, goodput excludes retransmissions, duplicate packets, protocol headers, and control messages. For link $(i,j)$, goodput can be defined as
\begin{equation}
    GP_{ij} =
    \frac{D^{useful}_{ij}}{\Delta t},
\end{equation}
where $D^{useful}_{ij}$ is the successfully delivered useful application data.

The system-level goodput is
\begin{equation}
    GP_{\mathcal{S}} =
    \frac{
    \sum_{(i,j)\in \mathcal{L}} D^{useful}_{ij}
    }
    {\Delta t}.
\end{equation}

Goodput efficiency can be defined as
\begin{equation}
    GPE_{\mathcal{S}} =
    \frac{GP_{\mathcal{S}}}{TH^{net}_{\mathcal{S}}}.
\end{equation}

A higher value of $GPE_{\mathcal{S}}$ indicates that most transmitted data are useful application-level data. The minimal acquisition \textbf{scope} is \caseScopeNode{} because useful application data can be measured locally at the receiving endpoint.
% Acquisition phase
The acquisition occurs during the \casePhaseOperational{} \textbf{phase} in order to evaluate effective communication performance during system execution.
% Acquisition method
The acquisition \textbf{method} is \caseAcquisitionCustomInstrumentation{} because the application must define which transmitted data are considered useful, requiring application-level instrumentation.

\subsection{Response Time}

Network response time measures the time between sending a request and receiving the corresponding response. It is useful for client-server, cloud, edge, and service-oriented DCCS applications. For a request generated by node $i$ and served by node $j$, response time can be expressed as
\begin{equation}
    RT^{net}_{ij} =
    t^{response}_{i} - t^{request}_{i}.
\end{equation}

It can be decomposed as
\begin{equation}
    RT^{net}_{ij} =
    L^{req}_{ij}
    +
    ST_j
    +
    L^{res}_{ji},
\end{equation}
where $L^{req}_{ij}$ is request transmission latency, $ST_j$ is service time at node $j$, and $L^{res}_{ji}$ is response transmission latency.

The average network response time is
\begin{equation}
    \overline{RT}^{net}_{\mathcal{S}} =
    \frac{1}{|\mathcal{R}|}
    \sum_{r\in \mathcal{R}} RT^{net}_{r},
\end{equation}
where $\mathcal{R}$ is the set of completed network requests. The minimal acquisition \textbf{scope} is \caseScopeNode{} because the elapsed time between sending a request and receiving the corresponding response can be measured locally by the client.
% Acquisition phase
The acquisition occurs during the \casePhaseOperational{} \textbf{phase} in order to detect communication delays during system execution.
% Acquisition method
The acquisition \textbf{method} is \caseAcquisitionStandardTelemetry{} because request and response timings are routinely exposed by standard networking and distributed tracing instrumentation.

\subsection{Service Time}

Service time measures the time spent by a network or computing node to process a request after receiving it. In DCCS, service time may include protocol handling, packet inspection, routing decision, data aggregation, task offloading decision, and local computation.

For a request $r$ served by node $j$, service time is
\begin{equation}
    ST_{j}^{r} =
    t^{complete}_{j,r} - t^{arrival}_{j,r},
\end{equation}
where $t^{arrival}_{j,r}$ is the time when request $r$ arrives at node $j$, and $t^{complete}_{j,r}$ is the time when node $j$ completes the service.

The average service time of node $j$ is
\begin{equation}
    \overline{ST}_{j} =
    \frac{1}{|\mathbf{R}_{j}|}
    \sum_{r\in \mathbf{R}_{j}} ST_{j}^{r},
\end{equation}
where $\mathbf{R}_{j}$ is the set of requests served by node $j$.

The system-level average service time is
\begin{equation}
    \overline{ST}_{\mathcal{S}} =
    \frac{1}{\zeta}
    \sum_{j\in \mathcal{N}}
    \overline{ST}_{j}.
\end{equation}
The minimal acquisition \textbf{scope} is \caseScopeNode{} because the arrival and completion times of a request are both measured locally at the serving node.
% Acquisition phase
The acquisition occurs during the \casePhaseOperational{} \textbf{phase} in order to detect processing delays during system execution.
% Acquisition method
The acquisition \textbf{method} is \caseAcquisitionStandardTelemetry{} because request processing durations are routinely exposed by standard tracing and monitoring instrumentation.

\subsection{Queueing Delay}

Queueing delay measures the time a packet or request spends waiting in a buffer before transmission or processing. It is an important metric because congestion in edge, fog, or gateway nodes may increase the total response time.

For packet $p$ waiting at node $j$, queueing delay is
\begin{equation}
    QD_{j}^{p} =
    t^{service}_{j,p} - t^{arrival}_{j,p},
\end{equation}
where $t^{arrival}_{j,p}$ is the packet arrival time and $t^{service}_{j,p}$ is the time when service begins.

The average queueing delay at node $j$ is
\begin{equation}
    \overline{QD}_{j} =
    \frac{1}{P_j}
    \sum_{p=1}^{P_j}
    QD_{j}^{p},
\end{equation}
where $P_j$ is the number of packets arriving at node $j$. The minimal acquisition \textbf{scope} is \caseScopeNode{} because packet arrival and service start times are both measured locally at the queue.
% Acquisition phase
The acquisition occurs during the \casePhaseOperational{} \textbf{phase} in order to detect congestion during system execution.
% Acquisition method
The acquisition \textbf{method} is \caseAcquisitionStandardTelemetry{} because queue occupancy and waiting times are routinely exposed by OS, network devices, and tracing instrumentation.

\subsection{Network Congestion Level}

Network congestion occurs when the traffic demand exceeds the available transmission or processing capacity. A simple congestion level for link $(i,j)$ can be defined using bandwidth utilization and queue occupancy:
\begin{equation}
    CG_{ij} =
    a U^{bw}_{ij}
    +
    (1-a) U^{buf}_{j},
\end{equation}
where $U^{bw}_{ij}$ is bandwidth utilization, $U^{buf}_{j}$ is buffer utilization at node $j$, and $0\leq a \leq 1$.

The system-level congestion level is
\begin{equation}
    CG_{\mathcal{S}} =
    \frac{1}{|\mathcal{L}|}
    \sum_{(i,j)\in \mathcal{L}}
    CG_{ij}.
\end{equation}

A congestion-aware packet drop probability can also be defined as
\begin{equation}
    P^{drop}_{j} =
    \frac{P^{drop}_{j}}{P^{arr}_{j}},
\end{equation}
where $P^{drop}_{j}$ is the number of packets dropped at node $j$ and $P^{arr}_{j}$ is the number of packets arriving at node $j$. The minimal acquisition \textbf{scope} is \caseScopeNode{} because it is computed from bandwidth utilization and buffer occupancy, both of which can be measured locally, as previously mentioned.
% Acquisition phase
The acquisition occurs during the \casePhaseOperational{} \textbf{phase} in order to detect congestion during system execution.
% Acquisition method
The acquisition \textbf{method} is \caseAcquisitionStandardTelemetry{} because bandwidth, queue occupancy, and packet drop counters are routinely exposed by network devices and OS.

\subsection{Network Availability}
Network availability measures the probability that a communication link, path, or network service is operational. For a communication link $(i,j)$, availability can be expressed as
\begin{equation}
    A^{net}_{ij} =
    \frac{MTTF_{ij}}{MTTF_{ij}+MTTR_{ij}},
\end{equation}
where $MTTF_{ij}$ is the mean time to failure and $MTTR_{ij}$ is the mean time to repair of link $(i,j)$.

The average network availability of the DCCS is
\begin{equation}
    A^{net}_{\mathcal{S}} =
    \frac{1}{|\mathcal{L}|}
    \sum_{(i,j)\in \mathcal{L}}
    A^{net}_{ij}.
\end{equation}

For a multi-hop path $P_{sd}$, path availability can be computed as
\begin{equation}
    A^{path}_{sd} =
    \prod_{(i,j)\in P_{sd}}
    A^{net}_{ij}.
\end{equation}

If multiple redundant paths $\mathcal{P}_{sd}$ are available between source $s$ and destination $d$, the redundant path availability is
\begin{equation}
    A^{red}_{sd} =
    1 -
    \prod_{P_{sd}\in \mathcal{P}_{sd}}
    \left(1-A^{path}_{sd}\right).
\end{equation}
The minimal acquisition \textbf{scope} is \caseScopeNode{} because link or service availability can be monitored locally using heartbeat or connectivity checks through pings.
% Acquisition phase
The acquisition occurs during the \casePhaseOperational{} \textbf{phase} in order to detect communication outages during system execution.
% Acquisition method
The acquisition \textbf{method} is \caseAcquisitionStandardTelemetry{} because availability is routinely monitored through standard heartbeat, keep-alive, and network monitoring mechanisms.

\subsection{Network Reliability}

Network reliability measures the probability that a communication link or path successfully operates without failure for a given time duration. If $\lambda_{ij}$ is the failure rate of link $(i,j)$, link reliability over time $t$ can be expressed as
\begin{equation}
    R^{net}_{ij}(t) =
    e^{-\lambda_{ij}t}.
\end{equation}

The reliability of a communication path $P_{sd}$ is
\begin{equation}
    R^{path}_{sd}(t) =
    \prod_{(i,j)\in P_{sd}}
    R^{net}_{ij}(t).
\end{equation}

For redundant communication paths, reliability can be improved as
\begin{equation}
    R^{red}_{sd}(t) =
    1 -
    \prod_{P_{sd}\in \mathcal{P}_{sd}}
    \left(1-R^{path}_{sd}(t)\right).
\end{equation}
The minimal acquisition \textbf{scope} is \caseScopeNode{} because link failures can be monitored locally to estimate the failure rate and reliability.
% Acquisition phase
The acquisition occurs during the \casePhaseOperational{} \textbf{phase} in order to detect communication failures during system execution.
% Acquisition method
The acquisition \textbf{method} is \caseAcquisitionStandardTelemetry{} because failure events and uptime are routinely exposed by standard network monitoring mechanisms.

\subsection{Handover Delay}

Handover delay is important in DCCS when mobile devices $\mathcal{M}$ move between different access points, edge nodes, or fog nodes \cite{salvatori2026dualgraphmultiagentreinforcementlearning, 11316643}. It measures the interruption time during mobility-driven service migration or network reconnection.

For mobile node $M_i$, handover delay can be defined as
\begin{equation}
    HD_{M_i} =
    t^{new}_{connect} - t^{old}_{disconnect},
\end{equation}
where $t^{old}_{disconnect}$ is the time when $M_i$ disconnects from the previous node, and $t^{new}_{connect}$ is the time when it connects to the new node.

The average handover delay for all mobile devices is
\begin{equation}
    \overline{HD}_{\mathcal{M}} =
    \frac{1}{m}
    \sum_{M_i\in \mathcal{M}}
    HD_{M_i}.
\end{equation}

A handover failure rate can be defined as
\begin{equation}
    HFR =
    \frac{H^{fail}}{H^{total}},
\end{equation}
where $H^{fail}$ is the number of failed handovers and $H^{total}$ is the total number of handover attempts. The minimal acquisition \textbf{scope} is \caseScopeNode{}* because the disconnection and reconnection times can be measured locally by the mobile device. Without instrumentation on the mobile device, telemetry from multiple nodes would be required.
% Acquisition phase
The acquisition occurs during the \casePhaseOperational{} \textbf{phase} in order to detect mobility-related communication interruptions during system execution.
% Acquisition method
The acquisition \textbf{method} is \caseAcquisitionStandardTelemetry{} because handover events are routinely exposed by the OS through network interface state changes and connectivity events.

\subsection{Communication Energy Consumption}

Energy consumption during communications is a metric that measures the energy spent for transmitting, receiving, and listening to network traffic \cite{feng2012survey,10944615}. This metric is essential for battery-powered sensor, IoT, edge, and mobile devices, and continuously exchange control signals and messages among devices over the network. For node $i$, communication energy consumption can be expressed as
\begin{equation}
    E^{comm}_{i} =
    E^{tx}_{i} + E^{rx}_{i} + E^{idle}_{i},
\end{equation}
where $E^{tx}_{i}$ is transmission energy, $E^{rx}_{i}$ is reception energy, and $E^{idle}_{i}$ is idle listening energy.

For data transmission from node $i$ to node $j$, transmission energy can be approximated as
\begin{equation}
    E^{tx}_{ij} =
    P^{tx}_{i} \times T^{tx}_{ij},
\end{equation}
where $P^{tx}_{i}$ is the transmission power of node $i$ and $T^{tx}_{ij}$ is the transmission duration.

If $d_{ij}$ amount of data is transmitted over bandwidth $B_{ij}$, then
\begin{equation}
    T^{tx}_{ij} =
    \frac{d_{ij}}{B_{ij}}.
\end{equation}

The total communication energy consumption of the DCCS is
\begin{equation}
    E^{comm}_{\mathcal{S}} =
    \sum_{i\in \mathcal{N}^{all}}
    E^{comm}_{i}.
\end{equation}
The minimal acquisition \textbf{scope} is \caseScopeNode{} because communication energy can be measured or estimated independently for each node.
% Acquisition phase
The acquisition occurs during the \casePhaseOperational{} \textbf{phase} in order to monitor the communication energy consumption of battery-powered devices during system execution.
% Acquisition method
The acquisition \textbf{method} is \caseAcquisitionExperimentalInstrumentation{} because communication energy is not directly exposed by standard telemetry and typically requires hardware power measurements or energy estimation models.
% \subsection{Communication Energy Efficiency}

Communication energy efficiency measures the amount of successfully delivered data per unit communication energy. It can be defined as
\begin{equation}
    CEE_{\mathcal{S}} =
    \frac{
    \sum_{(i,j)\in \mathcal{L}} D^{rx}_{ij}
    }
    {E^{comm}_{\mathcal{S}}}.
\end{equation}

For useful application-level data, goodput-based communication energy efficiency is
\begin{equation}
    GCEE_{\mathcal{S}} =
    \frac{
    \sum_{(i,j)\in \mathcal{L}} D^{useful}_{ij}
    }
    {E^{comm}_{\mathcal{S}}}.
\end{equation}

A higher value indicates that the network delivers more useful data with less energy consumption. The minimal acquisition \textbf{scope} is \caseScopeNode{} because both the delivered data and communication energy can be measured or estimated independently for each node.
% Acquisition phase
The acquisition occurs during the \casePhaseExperimental{} \textbf{phase} in order to evaluate the energy efficiency of communication mechanisms and network configurations.
% Acquisition method
The acquisition \textbf{method} is \caseAcquisitionExperimentalInstrumentation{} because it depends on communication energy measurements, which typically require hardware power measurements or energy estimation models.

\subsection{Connection Density}

Connection density measures measures the proportion of direct communication links among the nodes of the system. It is useful for understanding the connectivity structure of the system.  The maximum possible number of directed links among $n$ devices is $n(n-1)$. Therefore, directed connection density can be defined as
\begin{equation}
    CD_{\mathcal{S}} =
    \frac{
    \sum_{i=1}^{n}
    \sum_{\substack{j=1 \\ j\neq i}}^{n}
    \mathbb{C}_{ij}
    }
    {n(n-1)}.
\end{equation}

For undirected communication links, the connection density is
\begin{equation}
    CD_{\mathcal{S}} =
    \frac{
    2|\mathcal{L}|
    }
    {n(n-1)}.
\end{equation}

A very low value may indicate poor connectivity, while a very high value may increase maintenance complexity and network management overhead. The minimal acquisition \textbf{scope} is \caseScopeSystem{} because the complete network topology is required to determine the total number of nodes and communication links.
% Acquisition phase
The acquisition occurs during the \casePhaseOperational{} \textbf{phase} in order to monitor changes in the communication topology during system execution.
% Acquisition method
The acquisition \textbf{method} is \caseAcquisitionStandardTelemetry{} because network topology is routinely exposed through kernel networking interfaces (e.g., Netlink) and topology discovery protocols.

\subsection{Path Stretch}

Path stretch measures how much longer the selected communication path is compared with the shortest possible path. For a source-destination pair $(s,d)$, it is defined as
\begin{equation}
    PS_{sd} =
    \frac{Cost(P_{sd}^{selected})}
    {Cost(P_{sd}^{shortest})},
\end{equation}
where $Cost(P)$ may represent hop count, latency, energy, or routing cost.

If latency is used as the cost, then
\begin{equation}
    PS^{lat}_{sd} =
    \frac{L(P_{sd}^{selected})}
    {L(P_{sd}^{shortest})}.
\end{equation}

A value close to $1$ indicates efficient routing, while a larger value indicates suboptimal path selection. The minimal acquisition \textbf{scope} is \caseScopeSystem{} because evaluating the selected path against the shortest possible path for a source-destination pair requires knowledge of the complete communication graph.
% Acquisition phase
The acquisition occurs during the \casePhaseOperational{} \textbf{phase} in order to evaluate routing efficiency during system execution.
% Acquisition method
The acquisition \textbf{method} is \caseAcquisitionStandardTelemetry{} because the communication graph can be obtained from kernel networking information (e.g., Netlink) or reconstructed from distributed traces.

\subsection{Network Cost}

Network cost measures the operational or monetary cost of communication. This can include data transfer cost, network usage cost, energy cost, and service provider cost. For a link $(i,j)$, communication cost can be represented as
\begin{equation}
    Cost^{net}_{ij} =
    \kappa_{ij} D^{tx}_{ij},
\end{equation}
where $\kappa_{ij}$ is the cost per unit data transmitted over link $(i,j)$ and $D^{tx}_{ij}$ is the transmitted data volume.

The total network cost of DCCS is
\begin{equation}
    Cost^{net}_{\mathcal{S}} =
    \sum_{(i,j)\in \mathcal{L}}
    Cost^{net}_{ij}.
\end{equation}

Cost per successfully delivered data unit can be defined as
\begin{equation}
    CPD =
    \frac{
    Cost^{net}_{\mathcal{S}}
    }
    {
    \sum_{(i,j)\in \mathcal{L}} D^{rx}_{ij}
    }.
\end{equation}
The minimal acquisition \textbf{scope} is \caseScopeNode{} because transmitted data can be measured independently for each node.
% Acquisition phase
The acquisition occurs during the \casePhaseExperimental{} \textbf{phase} in order to evaluate communication strategies and deployment configurations. Although it can also be used by cost-aware orchestration systems at runtime, this remains application-dependent.
% Acquisition method
The acquisition \textbf{method} is \caseAcquisitionExperimentalInstrumentation{} because communication cost depends on external pricing or energy models rather than standard telemetry, and there is no universally accepted method to derive them.

\section{Application or User-level Performance Measurement}\label{sec:ApplicationLevel}

Application or user-level performance measurement evaluates how well a DCCS satisfies application requirements and user expectations. While computing-level metrics focus on resource execution and network-level metrics focus on communication performance, application-level metrics measure the quality perceived by applications, users, and service providers. These metrics include task success, cost, time and space complexity, accuracy, availability, error rate, flexibility, user satisfaction, service-level objective compliance, and learnability.  Further, table \ref{tab:application_metrics} summarizes the acquisition requirements associated with each application-level metric. %The following paragraphs explain the rationale behind each classification and discuss the technical considerations required for metric extraction.

\begin{table}[t]
\centering
\caption{Acquisition requirements for application or user-level metrics.}
\label{tab:application_metrics}
\begin{tabular}{|l|c|c|c|}
\hline
\textbf{Metric} & \textbf{Scope} & \textbf{Phase} & \textbf{Method} \\
\hline
\textbf{Task Success Ratio} & \caseScopeApp{} & \casePhaseOperational{} & \caseAcquisitionCustomInstrumentation{} \\
\hline
\textbf{Application Cost} & \caseScopeApp{} & \casePhaseExperimental{}* & \caseAcquisitionExperimentalInstrumentation{} \\
\hline
\textbf{Time \& Space Complexity} & \caseScopeApp{} & \casePhaseExperimental{} & \caseAcquisitionStandardTelemetry{} \\
\hline
\textbf{Accuracy} & \caseScopeNode{} & \casePhaseExperimental{} & \caseAcquisitionExperimentalInstrumentation{} \\
\hline
\textbf{Availability} & \caseScopeApp{} & \casePhaseOperational{} & \caseAcquisitionStandardTelemetry{} \\
\hline
\textbf{SLO and SLA Compliance} & \caseScopeApp{} & \casePhaseOperational{} & \caseAcquisitionExperimentalInstrumentation{} \\
\hline
\textbf{Error Rate} & \caseScopeApp{} & \casePhaseOperational{} & \caseAcquisitionStandardTelemetry{}* \\
\hline
\textbf{Efficiency Metrics} & \caseScopeApp{} & \casePhaseOperational{} & \caseAcquisitionCustomInstrumentation{} \\
\hline
\textbf{User Satisfaction Metrics} & \caseScopeNode{} & \casePhaseExperimental{} & \caseAcquisitionExperimentalInstrumentation{} \\
\hline
\textbf{Learnability Metrics} & \caseScopeNode{} & \casePhaseExperimental{} & \caseAcquisitionCustomInstrumentation{} \\
\hline
\textbf{Flexibility} & \caseScopeSystem{} & \casePhaseOperational{} & \caseAcquisitionCustomInstrumentation{} \\
\hline
\textbf{Application Portability} & \caseScopeSystem{} & \casePhaseExperimental{} & \caseAcquisitionCustomInstrumentation{} \\
\hline
\textbf{Application Robustness} & \caseScopeApp{} & \casePhaseExperimental{} & \caseAcquisitionCustomInstrumentation{}* \\
\hline
\end{tabular}
\end{table}

Let $\mathcal{A}=\{A_1,A_2,\ldots,A_a\}$ denote the set of applications deployed over $\mathcal{S}$. Each application $A_l \in \mathcal{A}$ consists of a set of tasks $\mathcal{T}_{A_l}=\{T_1,T_2,\ldots,T_q\}$. A task $T_i$ (similar to Eq. (\ref{eqTx})) is represented as
\begin{equation}
    T_i = \langle \omega_i, d_i, \tau_i, \delta_i, \gamma_i  \rangle,
\end{equation}
where $\omega_i$ is the computational workload, $d_i$ is the input data size, $\tau_i$ is the task arrival time, $\delta_i$ is the deadline, and $\gamma_i $ is the expected application-level quality requirement. Let $\mathcal{T}^{c}_{A_l}$ denote the set of completed tasks, $\mathcal{T}^{f}_{A_l}$ denote the set of failed tasks, and $\mathcal{T}^{s}_{A_l}$ denote the set of successfully completed tasks for application $A_l$.

\subsection{Task Success Metrics}

Task success measures whether an application task is completed correctly and within the required constraints. In DCCS, a task may be considered successful only if it satisfies functional correctness, deadline, quality, and resource constraints. The task success ratio of application $A_l$ is defined as
\begin{equation}
    TSR_{A_l} =
    \frac{|\mathcal{T}^{s}_{A_l}|}
    {|\mathcal{T}_{A_l}|}.
\end{equation}

For deadline-constrained applications, deadline-aware task success ratio is defined as
\begin{equation}
    DTSR_{A_l} =
    \frac{|\mathcal{T}^{s,\delta}_{A_l}|}
    {|\mathcal{T}_{A_l}|},
\end{equation}
where $\mathcal{T}^{s,\delta}_{A_l}$ is the set of tasks that are successfully completed before their deadlines.

If a task must satisfy both correctness and deadline constraints, then
\begin{equation}
    T_i^{s} =
    \begin{cases}
    1, & \text{if } Correct(T_i)=1 \text{ and } RT_i \leq \delta_i,\\
    0, & \text{otherwise}.
    \end{cases}
\end{equation}

The application-level task success ratio can then be written as
\begin{equation}
    TSR_{A_l} =
    \frac{1}{|\mathcal{T}_{A_l}|}
    \sum_{T_i \in \mathcal{T}_{A_l}} T_i^{s}.
\end{equation}

For DCCS applications where output quality is important, a quality-aware task success ratio can be defined as
\begin{equation}
    QTSR_{A_l} =
    \frac{1}{|\mathcal{T}_{A_l}|}
    \sum_{T_i \in \mathcal{T}_{A_l}}
    \mathbbm{1}(RT_i \leq \delta_i \land Q_i \geq \gamma_i ),
\end{equation}
where $Q_i$ is the achieved output quality, $\gamma_i $ is the minimum required quality, and $\mathbbm{1}(\cdot)$ is an indicator function. The minimal acquisition \textbf{scope} is \caseScopeApp{} because $TSR_{A_l}$ is defined over all tasks belonging to an application, which may execute on multiple nodes. However, it can be collected from a single node when the application is entirely executed there.
% Acquisition phase
The acquisition occurs during the \casePhaseOperational{} \textbf{phase} in order to monitor task execution during system operation.
% Acquisition method
The acquisition \textbf{method} is \caseAcquisitionCustomInstrumentation{} because the application must define what constitutes a successfully completed task, such as satisfying correctness, deadline, or quality constraints.

\subsection{Cost}

Application-level cost measures the monetary or operational cost required to execute an application over the DCCS \cite{sedlak_service_2026,loven2026realtimeaiserviceeconomy}. The total cost of application $A_l$ can include computing cost, communication cost, storage cost, energy cost, and migration cost:
\begin{equation}
    Cost_{A_l} =
    Cost^{comp}_{A_l}
    +
    Cost^{comm}_{A_l}
    +
    Cost^{sto}_{A_l}
    +
    Cost^{eng}_{A_l}
    +
    Cost^{mig}_{A_l}.
\end{equation}

The computing cost can be expressed as
\begin{equation}
    Cost^{comp}_{A_l} =
    \sum_{T_i \in \mathcal{T}_{A_l}}
    \sum_{j \in \mathcal{N}}
    x_{ij} \kappa^{comp}_{j} ET_{ij},
\end{equation}
where $x_{ij}$ indicates whether task $T_i$ is assigned to computing node $j$, $\kappa^{comp}_{j}$ is the computing cost per unit time of node $j$, and $ET_{ij}$ is the execution time of task $T_i$ on node $j$.

The communication cost is
\begin{equation}
    Cost^{comm}_{A_l} =
    \sum_{(i,j)\in \mathcal{L}_{A_l}}
    \kappa^{comm}_{ij} D^{tx}_{ij},
\end{equation}
where $\kappa^{comm}_{ij}$ is the communication cost per unit data over link $(i,j)$, and $D^{tx}_{ij}$ is the transmitted data volume.

The cost per successful task is defined as
\begin{equation}
    CPST_{A_l} =
    \frac{Cost_{A_l}}
    {|\mathcal{T}^{s}_{A_l}|}.
\end{equation}

The cost efficiency of application $A_l$ is
\begin{equation}
    CE_{A_l} =
    \frac{|\mathcal{T}^{s}_{A_l}|}
    {Cost_{A_l}}.
\end{equation}

For quality-aware applications, the cost-quality efficiency can be expressed as
\begin{equation}
    CQE_{A_l} =
    \frac{\sum_{T_i \in \mathcal{T}^{s}_{A_l}} Q_i}
    {Cost_{A_l}}.
\end{equation}
The minimal acquisition \textbf{scope} is \caseScopeApp{} because the metric aggregates the execution cost of all resources involved in the application, which may span multiple nodes.
% Acquisition phase
The acquisition occurs during the \casePhaseExperimental{}* \textbf{phase} because application cost is primarily used to evaluate deployment strategies and resource allocation. Although it can also be used by cost-aware orchestration systems at runtime, this remains application-dependent.
% Acquisition method
The acquisition \textbf{method} is \caseAcquisitionExperimentalInstrumentation{} because it depends on external pricing or energy models rather than standard telemetry, and there is no universally accepted method to derive it.

\subsection{Time and Space Complexity}

Time and space complexity describe the growth of application execution time and memory requirements with respect to input size. This traditional metrics not only applicable user level but also influence the systems level.  Let $n_i$ denote the input size of task $T_i$. The theoretical time complexity of the algorithm used by application $A_l$ can be represented as
\begin{equation}
    TC_{A_l}(n_i) = O(g(n_i)),
\end{equation}
where $g(n_i)$ represents the growth function of execution time.

The observed execution time of application $A_l$ over DCCS is
\begin{equation}
    OET_{A_l} =
    \sum_{T_i \in \mathcal{T}_{A_l}}
    \sum_{j \in \mathcal{N}}
    x_{ij} ET_{ij}.
\end{equation}

In DCCS, the practical application completion time also includes communication and coordination overhead:
\begin{equation}
    ACT_{A_l} =
    \sum_{T_i \in \mathcal{T}_{A_l}}
    \left(
    ET_i + CT_i + QT_i + SY_i + MG_i
    \right),
\end{equation}
where $ET_i$ is execution time, $CT_i$ is communication time, $QT_i$ is queueing time, $SY_i$ is synchronization time, and $MG_i$ is migration time.

The space complexity of application $A_l$ is represented as
\begin{equation}
    SC_{A_l}(n_i) = O(h(n_i)),
\end{equation}
where $h(n_i)$ represents the growth function of memory consumption.

The observed memory usage of application $A_l$ is
\begin{equation}
    OMU_{A_l} =
    \sum_{j \in \mathcal{N}}
    M^{used}_{j,A_l},
\end{equation}
where $M^{used}_{j,A_l}$ denotes the memory consumed by application $A_l$ on node $j$.

A continuum overhead ratio can be defined as
\begin{equation}
    COR_{A_l} =
    \frac{ACT_{A_l} - OET_{A_l}}
    {ACT_{A_l}}.
\end{equation}
A lower value of $COR_{A_l}$ indicates that less time is spent on communication, synchronization, queueing, and migration overhead. The minimal acquisition \textbf{scope} is \caseScopeApp{} because the observed execution time, memory usage, and continuum overhead are aggregated across all tasks and nodes belonging to the application.
% Acquisition phase
The acquisition occurs during the \casePhaseExperimental{} \textbf{phase} because these metrics are primarily used to evaluate scalability and identify the most efficient algorithms, deployments, or execution configurations by comparing different executions.
% Acquisition method
The acquisition \textbf{method} is \caseAcquisitionStandardTelemetry{} because execution times, memory usage, communication delays, synchronization events, and migration events are routinely exposed by OS, orchestrators, and distributed tracing frameworks.

\subsection{Accuracy}

Accuracy measures the correctness of application outputs. In DCCS, accuracy is important for applications such as classification, prediction, object detection, anomaly detection, decision support, and control. For classification-based applications, accuracy is defined as
\begin{equation}
    Acc =
    \frac{TP + TN}
    {TP + TN + FP + FN},
\end{equation}
where $TP$, $TN$, $FP$, and $FN$ denote true positives, true negatives, false positives, and false negatives, respectively.

Precision is defined as
\begin{equation}
    Precision =
    \frac{TP}{TP+FP},
\end{equation}
and recall is defined as
\begin{equation}
    Recall =
    \frac{TP}{TP+FN}.
\end{equation}

The F1-score is the harmonic mean of precision and recall:
\begin{equation}
    F1 =
    2 \times
    \frac{Precision \times Recall}
    {Precision + Recall}.
\end{equation}

For regression or prediction-based applications, mean absolute error can be used:
\begin{equation}
    MAE =
    \frac{1}{N}
    \sum_{i=1}^{N}
    |y_i-\hat{y}_i|,
\end{equation}
where $y_i$ is the actual value and $\hat{y}_i$ is the predicted value.

Mean squared error is defined as
\begin{equation}
    MSE =
    \frac{1}{N}
    \sum_{i=1}^{N}
    (y_i-\hat{y}_i)^2.
\end{equation}

Root mean squared error is
\begin{equation}
    RMSE =
    \sqrt{
    \frac{1}{N}
    \sum_{i=1}^{N}
    (y_i-\hat{y}_i)^2
    }.
\end{equation}

For DCCS, accuracy can also be measured under resource constraints. Therefore, resource-aware accuracy can be defined as
\begin{equation}
    RAcc_{A_l} =
    \frac{Acc_{A_l}}
    {w_1 RT_{A_l} + w_2 E_{A_l} + w_3 Cost_{A_l}},
\end{equation}
where $RT_{A_l}$ is response time, $E_{A_l}$ is energy consumption, $Cost_{A_l}$ is application cost, and $w_1$, $w_2$, and $w_3$ are weighting factors. The minimal acquisition \textbf{scope} is \caseScopeNode{} because this metric is evaluated on the prediction or inference nodes, and all the required information can be obtained from a single node when the inference is performed locally. Applications requiring distributed inference would instead require telemetry from multiple nodes.
% Acquisition phase
The acquisition occurs during the \casePhaseExperimental{} \textbf{phase} because evaluating accuracy requires comparing predictions against reference outputs, making it primarily suitable for validation, benchmarking, and model selection rather than continuous runtime monitoring.
% Acquisition method
The acquisition \textbf{method} is \caseAcquisitionExperimentalInstrumentation{} because it requires external ground truth or an oracle to determine whether the produced outputs are correct.

\subsection{Availability}

Application-level availability measures the probability that an application is accessible and operational when requested by users. It can be defined using uptime and downtime:
\begin{equation}
    A_{A_l} =
    \frac{T^{up}_{A_l}}
    {T^{up}_{A_l}+T^{down}_{A_l}},
\end{equation}
where $T^{up}_{A_l}$ is the total application uptime and $T^{down}_{A_l}$ is the total downtime.

Using mean time to failure and mean time to repair, availability can also be expressed as
\begin{equation}
    A_{A_l} =
    \frac{MTTF_{A_l}}
    {MTTF_{A_l}+MTTR_{A_l}}.
\end{equation}

For an application deployed across multiple continuum layers, application availability depends on the availability of required computing nodes, communication links, and software services. If application $A_l$ requires a set of nodes $\mathcal{N}_{A_l}$ and links $\mathcal{L}_{A_l}$, then a simple serial dependency availability model can be expressed as
\begin{equation}
    A^{serial}_{A_l} =
    \prod_{j \in \mathcal{N}_{A_l}} A_j
    \prod_{(i,j)\in \mathcal{L}_{A_l}} A^{net}_{ij}.
\end{equation}

If the application can use redundant nodes or alternative paths, availability can be improved as
\begin{equation}
    A^{red}_{A_l} =
    1 -
    \prod_{r \in \mathcal{R}_{A_l}}
    (1-A_r),
\end{equation}
where $\mathcal{R}_{A_l}$ represents redundant application instances, alternative execution paths, or service replicas. The minimal acquisition \textbf{scope} is \caseScopeApp{} because, although an application may be accessible through a single entry point, the metric evaluates whether the entire application remains operational. This generally requires considering the availability of all nodes involved in the application, or at least those required to execute the requested functionality.
% Acquisition phase
The acquisition occurs during the \casePhaseOperational{} \textbf{phase} in order to detect service interruptions and ensure that the application remains accessible during system execution.
% Acquisition method
The acquisition \textbf{method} is \caseAcquisitionStandardTelemetry{} because application availability is routinely monitored through health checks, heartbeat mechanisms, service status monitoring, and uptime records.

\subsection{SLO and SLA Compliance}

Service-level objectives (SLOs) and service-level agreements (SLAs) define expected service quality from the application or user perspective \cite{11298469}. Let $\mathcal{O}_{A_l}=\{O_1,O_2,\ldots,O_o\}$ denote the set of SLOs for application $A_l$. Each SLO $O_r$ can be represented as
\begin{equation}
    O_r = \langle m_r, \theta_r, \diamond_r \rangle,
\end{equation}
where $m_r$ is the measured metric, $\theta_r$ is the target threshold, and $\diamond_r \in \{\leq,\geq\}$ is the satisfaction condition.

The SLO satisfaction indicator is
\begin{equation}
    Sat(O_r) =
    \begin{cases}
    1, & \text{if } m_r \diamond_r \theta_r,\\
    0, & \text{otherwise}.
    \end{cases}
\end{equation}

The SLO compliance ratio of application $A_l$ is
\begin{equation}
    SLOCR_{A_l} =
    \frac{1}{|\mathcal{O}_{A_l}|}
    \sum_{O_r \in \mathcal{O}_{A_l}}
    Sat(O_r).
\end{equation}

For task-level SLOs, the compliance ratio can be defined as
\begin{equation}
    TSLOCR_{A_l} =
    \frac{1}{|\mathcal{T}_{A_l}|}
    \sum_{T_i \in \mathcal{T}_{A_l}}
    \mathbbm{1}
    \left(
    RT_i \leq \delta_i
    \land
    Q_i \geq \gamma_i 
    \land
    Cost_i \leq Cost_i^{max}
    \right).
\end{equation}

An SLA violation rate can be defined as
\begin{equation}
    SLAVR_{A_l} =
    1 - SLOCR_{A_l}.
\end{equation}
The minimal acquisition \textbf{scope} is \caseScopeApp{} because, although SLO and SLA compliance can be evaluated by a single monitoring component, that component requires telemetry from all application components involved in delivering the requested service. Therefore, the acquisition scope is transitively the application.
% Acquisition phase
The acquisition occurs during the \casePhaseOperational{} \textbf{phase} in order to continuously verify that the application satisfies its service-level objectives during execution. It can also support SLO-aware applications that dynamically adapt their configuration or resource allocation at runtime to maintain compliance.
% Acquisition method
The acquisition \textbf{method} is \caseAcquisitionExperimentalInstrumentation{} because SLO and SLA compliance require an external governance component that defines the objectives, collects telemetry from the monitored system, and evaluates whether the agreements are satisfied, as performed by frameworks such as Governify \cite{FRESNOARANDA2024100629}.

\subsection{Error Rate}

Error rate measures the fraction of application operations, tasks, requests, or user actions that result in incorrect or failed outcomes. The task error rate of application $A_l$ is defined as
\begin{equation}
    ER_{A_l} =
    \frac{|\mathcal{T}^{err}_{A_l}|}
    {|\mathcal{T}_{A_l}|},
\end{equation}
where $\mathcal{T}^{err}_{A_l}$ is the set of tasks that produce incorrect outputs or fail due to application-level errors.

If user actions are considered, user error rate is defined as
\begin{equation}
    UER =
    \frac{N^{user}_{err}}
    {N^{user}_{act}},
\end{equation}
where $N^{user}_{err}$ is the number of incorrect user actions and $N^{user}_{act}$ is the total number of user actions.

For request-based applications, request error rate can be expressed as
\begin{equation}
    RER_{A_l} =
    \frac{N^{failed}_{req}}
    {N^{total}_{req}}.
\end{equation}

The recoverable error ratio is
\begin{equation}
    RER^{rec}_{A_l} =
    \frac{N^{recovered}_{err}}
    {N^{total}_{err}},
\end{equation}
where $N^{recovered}_{err}$ is the number of errors successfully recovered by the system. The minimal acquisition \textbf{scope} is \caseScopeApp{} because the metric is defined over all tasks or requests belonging to the application, requiring the aggregation of errors across the application's execution.
% Acquisition phase
The acquisition occurs during the \casePhaseOperational{} \textbf{phase} in order to detect failures and monitor the application's health during execution.
% Acquisition method
The acquisition \textbf{method} is \caseAcquisitionStandardTelemetry{}* because failed requests and unsuccessful task executions are routinely exposed by logs and distributed tracing frameworks. However, application-specific definitions of incorrect outcomes may require \caseAcquisitionCustomInstrumentation{}.

\subsection{Efficiency Metrics}

Application efficiency measures how much useful output is produced relative to time, cost, energy, or resources. The time efficiency of application $A_l$ is
\begin{equation}
    TE_{A_l} =
    \frac{|\mathcal{T}^{s}_{A_l}|}
    {ACT_{A_l}}.
\end{equation}

Energy efficiency is defined as
\begin{equation}
    EE_{A_l} =
    \frac{|\mathcal{T}^{s}_{A_l}|}
    {E_{A_l}},
\end{equation}
where $E_{A_l}$ is the total energy consumed by application $A_l$.

Resource efficiency can be expressed as
\begin{equation}
    RE_{A_l} =
    \frac{|\mathcal{T}^{s}_{A_l}|}
    {\sum_{j\in \mathcal{N}} U^{res}_{j,A_l}},
\end{equation}
where $U^{res}_{j,A_l}$ denotes the amount of resources consumed by application $A_l$ on node $j$.

A combined application efficiency score can be defined as
\begin{equation}
    AE_{A_l} =
    yb_1 \widehat{TE}_{A_l}
    +
    yb_2 \widehat{CE}_{A_l}
    +
    yb_3 \widehat{EE}_{A_l}
    +
    yb_4 \widehat{RE}_{A_l},
\end{equation}
where $\widehat{TE}_{A_l}$, $\widehat{CE}_{A_l}$, $\widehat{EE}_{A_l}$, and $\widehat{RE}_{A_l}$ are normalized time, cost, energy, and resource efficiency values, respectively, and  $yb_1+yb_2+yb_3+yb_4=1$.
The minimal acquisition \textbf{scope} is \caseScopeApp{} because the metrics are computed from the successful tasks and resource consumption of the application as a whole.
% Acquisition phase
The acquisition occurs during the \casePhaseOperational{} \textbf{phase} in order to monitor the application's efficiency and detect performance degradation during execution.
% Acquisition method
The acquisition \textbf{method} is \caseAcquisitionCustomInstrumentation{} because the metrics depend on the application-specific definition of successful task completion, while the remaining quantities, such as execution time, energy, cost, and resource usage, can be obtained through standard telemetry.

\subsection{User Satisfaction Metrics}

User satisfaction measures the perceived quality of the application by end users. It can be collected through feedback scores, ratings, surveys, or interaction logs. Let $s_u$ denote the satisfaction score given by user $u$, where $s_u \in [1,5]$ or $s_u \in [1,10]$. The average user satisfaction score is
\begin{equation}
    USS =
    \frac{1}{|\mathcal{U}|}
    \sum_{u\in \mathcal{U}} s_u,
\end{equation}
where $\mathcal{U}$ is the set of users.

A normalized user satisfaction score can be expressed as
\begin{equation}
    NUSS =
    \frac{USS - s^{min}}
    {s^{max}-s^{min}},
\end{equation}
where $s^{min}$ and $s^{max}$ are the minimum and maximum possible satisfaction scores.

For interactive DCCS applications, satisfaction may depend on latency, availability, accuracy, and error rate. Therefore, a quality-of-experience score can be defined as
\begin{equation}
    QoE_{A_l} =
    \eta_1 \widehat{Acc}_{A_l}
    +
    \eta_2 \widehat{A}_{A_l}
    +
    \eta_3 \widehat{TSR}_{A_l}
    -
    \eta_4 \widehat{RT}_{A_l}
    -
    \eta_5 \widehat{ER}_{A_l},
\end{equation}
where $\widehat{Acc}_{A_l}$, $\widehat{A}_{A_l}$, and $\widehat{TSR}_{A_l}$ are normalized positive indicators, while $\widehat{RT}_{A_l}$ and $\widehat{ER}_{A_l}$ are normalized negative indicators. The minimal acquisition \textbf{scope} is \caseScopeNode{} because user satisfaction is typically collected and stored by a single feedback or analytics component that acts as the source of truth. Therefore, instrumenting the application nodes themselves is not required to obtain the metric.
% Acquisition phase
The acquisition occurs during the \casePhaseExperimental{} \textbf{phase} because user satisfaction is primarily used to evaluate and compare application designs, configurations, or deployments from the users' perspective.
% Acquisition method
The acquisition \textbf{method} is \caseAcquisitionExperimentalInstrumentation{} because user satisfaction requires an external feedback collection mechanism, such as surveys, ratings, questionnaires, or interaction analysis, which must be specifically instrumented to expose these data.

\subsection{Learnability Metrics}

Learnability measures how easily users learn to use an application. It is particularly relevant for human-facing DCCS applications, dashboards, mobile interfaces, digital twins, and control systems. Let $TTF_u$ denote the time taken by user $u$ to complete a task correctly for the first time. The average time-to-first-success is
\begin{equation}
    ATFS =
    \frac{1}{|\mathcal{U}|}
    \sum_{u\in \mathcal{U}} TTF_u.
\end{equation}

The first-attempt success ratio is
\begin{equation}
    FASR =
    \frac{N^{first}_{success}}
    {N^{first}_{attempt}},
\end{equation}
where $N^{first}_{success}$ is the number of users who successfully complete the task on the first attempt, and $N^{first}_{attempt}$ is the total number of first attempts.

Learning improvement between two sessions can be defined as
\begin{equation}
    LI =
    \frac{T^{session1}_{avg} - T^{session2}_{avg}}
    {T^{session1}_{avg}},
\end{equation}
where $T^{session1}_{avg}$ and $T^{session2}_{avg}$ are average task completion times in the first and second sessions, respectively.

Error reduction due to learning can be expressed as
\begin{equation}
    ERL =
    \frac{ER^{session1} - ER^{session2}}
    {ER^{session1}}.
\end{equation}
The minimal acquisition \textbf{scope} is \caseScopeNode{} because user interaction events can typically be collected by a single frontend or analytics component without requiring instrumentation of the entire application.
% Acquisition phase
The acquisition occurs during the \casePhaseExperimental{} \textbf{phase} because learnability is primarily evaluated during usability studies or controlled deployments to assess and compare interface designs and user experience.
% Acquisition method
The acquisition \textbf{method} is \caseAcquisitionCustomInstrumentation{} because the application must record user-specific events such as first attempts, successful task completions, and session progression, which are not provided by standard telemetry frameworks.

\subsection{Flexibility}

Flexibility measures the ability of an application to adapt to changing users, workloads, deployment environments, devices, and quality requirements. In DCCS, flexibility is important because applications may need to move across cloud, fog, edge, mobile, and IoT layers.

The deployment flexibility of application $A_l$ can be defined as
\begin{equation}
    DF_{A_l} =
    \frac{|\mathcal{N}^{eligible}_{A_l}|}
    {|\mathcal{N}|},
\end{equation}
where $\mathcal{N}^{eligible}_{A_l}$ is the set of computing nodes capable of executing application $A_l$.

The migration flexibility can be defined as
\begin{equation}
    MF_{A_l} =
    \frac{|\mathcal{P}^{mig}_{A_l}|}
    {|\mathcal{P}^{all}_{A_l}|},
\end{equation}
where $\mathcal{P}^{mig}_{A_l}$ is the set of feasible migration paths and $\mathcal{P}^{all}_{A_l}$ is the set of all possible migration paths.

The configuration flexibility can be defined as
\begin{equation}
    CF_{A_l} =
    \frac{N^{valid}_{config}}
    {N^{total}_{config}},
\end{equation}
where $N^{valid}_{config}$ is the number of valid application configurations that satisfy minimum requirements, and $N^{total}_{config}$ is the total number of possible configurations.

An application flexibility score can be computed as
\begin{equation}
    FS_{A_l} =
    wi_1 DF_{A_l}
    +
    w_2 MF_{A_l}
    +
    w_3 CF_{A_l}
    +
    w_4 AF_{A_l},
\end{equation}
where $AF_{A_l}$ denotes adaptation flexibility, and $w_1+w_2+w_3+w_4=1$.
% \end{equation}
The minimal acquisition \textbf{scope} is \caseScopeSystem{} because evaluating deployment, migration, and configuration flexibility requires knowledge of all available computing nodes and resources in the DCCS to determine which are eligible for the application.
% Acquisition phase
The acquisition occurs during the \casePhaseOperational{} \textbf{phase} in order to support adaptation decisions as the set of available resources and deployment opportunities changes during system execution.
% Acquisition method
The acquisition \textbf{method} is \caseAcquisitionCustomInstrumentation{} because the application must define eligibility rules, valid configurations, and feasible migration paths, which are application-specific and cannot be inferred from standard telemetry alone.

\subsection{Application Portability}

Application portability measures the ability of an application to be deployed or migrated across heterogeneous continuum nodes without major modification. Let $\mathcal{P}_{A_l}$ denote the set of platforms that support application $A_l$, and $\mathcal{P}^{all}$ denote the set of all available platforms in $\mathcal{S}$. Portability can be defined as
\begin{equation}
    PRT_{A_l} =
    \frac{|\mathcal{P}_{A_l}|}
    {|\mathcal{P}^{all}|}.
\end{equation}

A dependency-aware portability metric can be defined as
\begin{equation}
    DPRT_{A_l} =
    1 -
    \frac{N^{unsatisfied}_{dep}}
    {N^{total}_{dep}},
\end{equation}
where $N^{unsatisfied}_{dep}$ is the number of unsatisfied software, hardware, or network dependencies, and $N^{total}_{dep}$ is the total number of dependencies. The minimal acquisition \textbf{scope} is \caseScopeSystem{} because portability must be evaluated against all available platforms and resources in the DCCS to determine where the application can be successfully deployed.
% Acquisition phase
The acquisition occurs during the \casePhaseExperimental{} \textbf{phase} because portability is primarily assessed when evaluating deployment strategies, target infrastructures, or application designs before deployment.
% Acquisition method
The acquisition \textbf{method} is \caseAcquisitionCustomInstrumentation{} because the application must define its platform compatibility and software, hardware, and network dependencies, which cannot be inferred from standard telemetry alone. 

\subsection{Application Robustness}

Application robustness measures the ability of an application to continue producing acceptable outputs under failures, uncertainty, workload changes, and resource variations. Let $Q_i^{normal}$ be the output quality of task $T_i$ under normal conditions and $Q_i^{stress}$ be the output quality under stressed conditions. Robustness can be defined as
\begin{equation}
    ROB_{A_l} =
    \frac{1}{|\mathcal{T}_{A_l}|}
    \sum_{T_i \in \mathcal{T}_{A_l}}
    \frac{Q_i^{stress}}{Q_i^{normal}}.
\end{equation}

For deadline-sensitive applications, robustness can also be measured using the degradation in deadline-aware task success:
\begin{equation}
    DROB_{A_l} =
    1 -
    \left|
    DTSR^{normal}_{A_l}
    -
    DTSR^{stress}_{A_l}
    \right|.
\end{equation}

A value close to $1$ indicates that the application maintains similar performance under normal and stressed conditions. The minimal acquisition \textbf{scope} is \caseScopeApp{} because robustness is evaluated over the quality degradation of tasks belonging to the application under stressed conditions.
% Acquisition phase
The acquisition occurs during the \casePhaseExperimental{} \textbf{phase} because robustness evaluation requires controlled experiments that introduce stress conditions and compare application behavior against normal executions to improve the design of more resilient applications.
% Acquisition method
The acquisition \textbf{method} is \caseAcquisitionCustomInstrumentation{}* because the application must define how task quality and acceptable outputs are measured under stress conditions. In some cases, robustness evaluation may require \caseAcquisitionExperimentalInstrumentation{} when it depends on external data sources, complex models, or previously unavailable measurements.

\section{Novel Metric for Changing DCC demands}\label{sec:novelmetrics}
Researchers have proposed a wide range of performance measures for computing systems over the years. Yet as these systems keep growing in scale and complexity, and applications place ever-changing demands on them, many of the established metrics no longer tell the full story. This gap calls for fresh evaluation approaches that can capture what actually matters in today's distributed environments.

\subsection{$CO_2$ emissions} GenAI and LLMs are among the most resource-hungry workloads in modern computing, and their environmental footprint reflects that. Training a state-of-the-art model can consume electricity on par with what several hundred homes use over an entire year, generating hundreds of tons of ($CO_2$) along the way. Once deployed, these models continue to draw substantial power for everyday inference — often more per query than a standard web search. The data centers hosting them also require considerable water for cooling, which can strain local supplies. At the current pace, data-center carbon emissions could triple by 2030 to roughly 2.5 billion tons a year. Manufacturing the GPUs and accelerators that make all of this possible only adds to the tally.

Now consider what happens when GenAI starts running not in a handful of data centers but on millions of IoT and edge devices scattered around the world (e.g., smart home assistants, wearable health monitors, industrial sensors, autonomous vehicles, and everything in between). The energy implications are striking. These devices typically operate where power and cooling are scarce (e.g., on a factory floor, on someone's wrist, or inside a moving car). Even a lightweight AI model needs meaningful processing power, which translates into faster battery drain, greater hardware wear, and more frequent maintenance. At scale, the combined energy demand of all those devices can strain electricity grids and noticeably increase global $CO_2$ output. Designing new system models for the computing continuum, therefore, demands that carbon-related metrics be built in from the start.

One concrete step in this direction comes from Vahdat et al.~\cite{vahdat2024new}, who proposed a goodput metric that captures carbon dioxide and equivalent emissions alongside traditional performance indicators. Their formulation draws on the Greenhouse Gas Protocol and ISO standards 14040 and 14044, lending it the rigour needed for consistent, cross-system comparisons. In DCC and IoT environments, where sustainability pressures are growing and device footprints add up quickly, such a metric fills an increasingly important gap. This section introduce various novel performance metrics according to the changes in computational demands due to AI and Table \ref{tab:changing_dcc_demands_metrics} summarizes their acquisition requirements. %The following paragraphs explain the rationale behind each classification and discuss the technical considerations required for metric extraction.

\begin{table}[t]
\centering
\caption{A summary of Acquisition requirements for novel performance metrics for DCC.}
\label{tab:changing_dcc_demands_metrics}
\begin{tabular}{|l|c|c|c|}
\hline
\textbf{Metric} & \textbf{Scope} & \textbf{Phase} & \textbf{Method} \\
\hline

\textbf{CO$_2$ Emissions} & \caseScopeNode{} & \casePhaseOperational{} & \caseAcquisitionExperimentalInstrumentation{} \\
\hline
\textbf{Heat Dissipation} & \caseScopeNode{} & \casePhaseOperational{} & \caseAcquisitionExperimentalInstrumentation{} \\
\hline

\textbf{Bottlenecks} & \caseScopeSystem{} & \casePhaseOperational{} & \caseAcquisitionStandardTelemetry{} \\
\hline
\textbf{Measure Observability} & \caseScopeSystem{} & \casePhaseExperimental{} & \caseAcquisitionExperimentalInstrumentation{} \\
\hline

\textbf{Adaptivity Quotient} & \caseScopeSystem{} & \casePhaseExperimental{} & \caseAcquisitionStandardTelemetry*{} \\
\hline

\textbf{Maintain Equilibrium} & \caseScopeSystem{} & \casePhaseExperimental{} & \caseAcquisitionExperimentalInstrumentation{} \\
\hline

\textbf{Data Locality Index} & \caseScopeMulti{} & \casePhaseOperational{} & \caseAcquisitionStandardTelemetry{} \\
\hline

\textbf{CC Fragmentation Index} & \caseScopeSystem{} & \casePhaseOperational{} & \caseAcquisitionStandardTelemetry{} \\
\hline

\textbf{Migration Stability Index} & \caseScopeSystem{} & \casePhaseOperational{} & \caseAcquisitionStandardTelemetry{} \\
\hline

\textbf{Adaptation Cost Efficiency} & \caseScopeMulti{} & \casePhaseExperimental{} & \caseAcquisitionExperimentalInstrumentation{} \\
\hline

\textbf{SLO Viol. Recovery Time} & \caseScopeApp{} & \casePhaseOperational{} & \caseAcquisitionExperimentalInstrumentation{} \\
\hline

\textbf{C. Efficiency per Useful Task} & \caseScopeApp{} & \casePhaseExperimental{} & \caseAcquisitionExperimentalInstrumentation{} \\
\hline

\textbf{Water Usage Efficiency} & \caseScopeApp{} & \casePhaseExperimental{} & \caseAcquisitionExperimentalInstrumentation{} \\
\hline

\textbf{Data Freshness } &\caseScopeMulti{} & \casePhaseOperational{} & \caseAcquisitionStandardTelemetry{} \\
\hline

\textbf{Trustworthiness Score} & \caseScopeNode{} & \casePhaseOperational{} & \caseAcquisitionExperimentalInstrumentation{} \\
\hline

\end{tabular}
\end{table}

\subsection{Heat Dissipation}
Heat dissipation sits alongside $CO_2$ emissions as a key indicator of a system's environmental footprint. Every active device turns some of its power into waste heat; how efficiently that heat is removed determines cooling costs, hardware reliability, and long-term
sustainability.

The standard starting point for modelling heat loss is \textit{Newton's law of cooling}~\cite{winterton1999newton}, which
ties the cooling rate to the temperature gap between device and
 \begin{equation}\label{eq_Newton}
    T(t) = T_{\text{e}} + \bigl(T_0 - T_{\text{e}}\bigr)\, e^{-kt},
\end{equation}

where $T(t)$ is the device temperature at time $t$; $T_{\text{e}}$ denotes the ambient (environmental) temperature of the location where the device is deployed; $T_0$ represents the initial temperature of the device; $k$ is a positive constant that depends on the properties of the device and the cooling medium (i.e., it determines the rate of heat loss); and $e$ denotes Euler's number (approximately 2.718), which is the base of the natural logarithm. Simple as it is, this model gives designers a practical tool for predicting temperature evolution and right-sizing cooling strategies across every layer of the continuum (i.e., from warehouse-scale data centers to palm-sized IoT sensors). 

The minimal acquisition \textbf{scope} is \caseScopeNode{} because the heat generated by each device can be measured or estimated locally and later aggregated to characterize the thermal footprint of the entire DCCS.
% Acquisition phase
The acquisition occurs during the \casePhaseOperational{} \textbf{phase} in order to monitor the thermal behavior of deployed devices, detect overheating conditions, and evaluate thermal efficiency over time.
% Acquisition method
The acquisition method is \caseAcquisitionExperimentalInstrumentation{} because heat dissipation can be estimated from experimental measurements of the temperature difference between the inside and outside of the enclosure, rather than from standard telemetry.

\subsection{Bottlenecks}
In DCC, resource usage is often uneven. Some devices may be heavily loaded with tasks while others remain underutilized. Similarly, energy consumption can be much higher on certain nodes, and bandwidth availability may fluctuate across different parts of the network. These conditions create potential bottlenecks that are often hidden when relying on average metrics, as such measures tend to smooth out the underlying imbalances. To effectively identify and measure these hidden bottlenecks, the \textbf{fairness index} provides a valuable tool for assessing how evenly resources are distributed across the system. \textit{Jain’s fairness index} is one of the most popular metric to measure it \cite{jain1999throughput}, and  shown in Eq.~(\ref{eq_Jain}).
\begin{equation}\label{eq_Jain}
    \mathcal{F}(x_1, x_2, \ldots, x_n) = \frac{\left(\sum_{i=1}^n x_i\right)^2}{n \sum_{i=1}^n x_i^2}
\end{equation}
where $x_i$ represents the resource metric for the $i^{th}$ node—such as CPU load, energy consumption, or bandwidth usage—and $n$ is the total number of nodes. The index produces a value between 0 and 1, where 1 indicates perfect no bottlenecks, and values closer to 0 indicate increasing bottlenecks. Several fairness index variants based on Jain’s approach have been proposed; \textit{Donta et al.} \cite{8882255} offers an alternate formulation. The minimal acquisition \textbf{scope} is \caseScopeSystem{} because bottlenecks are identified by comparing the resource utilization of all participating nodes.
% Acquisition phase
The acquisition occurs during the \casePhaseOperational{} \textbf{phase} in order to detect resource imbalances during execution and support load balancing, and resource management decisions.
% Acquisition method
The acquisition \textbf{method} is \caseAcquisitionStandardTelemetry{} because the required resource metrics, such as CPU utilization, memory usage, bandwidth consumption, and energy consumption, are routinely exposed by OS, orchestrators, and standard monitoring frameworks.

\subsection{Measure Observability} Measure observability in a DCCS goes beyond simply watching dashboards and it means understanding \emph{why} the system behaves the way it does. When the internal processes and interactions of a distributed system are transparent and explainable, operators can move from reactive firefighting to proactive tuning: monitoring in real time, diagnosing root causes across cloud, edge, and IoT layers, and continuously improving performance. In short, explainability turns raw telemetry into actionable insight, keeping complex, heterogeneous, and increasingly autonomous systems accountable and adaptable.

To quantify this property, we define an observability score that combines two complementary ideas: how well each individual node can explain its own decisions, and how consistently cause-and-effect relationships flow between nodes. Formally:

\begin{equation}\label{eq_Observe}
\mathcal{E} = \frac{ \sum_{i=1}^{N} \gamma_i \, E_{\text{local}}^i }
{ N } \cdot \left( 1 - \frac{ \sum_{i \neq j} |CI_{ij} - CI_{ji}| }
{ \sum_{i,j} CI_{ij}} \right),
\end{equation}

where $N$ is the total number of nodes, $E_{\text{local}}^i$ is the local explainability score of node~$i$, and $\gamma_i$ is its
associated weight. $CI_{ij}$ captures the causal influence that node~$i$ exerts on node~$j$, estimated through techniques such as Granger causality or other causal inference methods~\cite{pujol2024causality}. The first factor gives the weighted-average explainability across the system. The second factor penalises the score whenever inter-node causal relationships are asymmetric or poorly understood which reflects the intuition that a truly observable system should exhibit clear, consistent cause-and-effect links between its components.

The minimal acquisition \textbf{scope} is \caseScopeSystem{} because observability depends on both the explainability of individual nodes and the causal relationships between components throughout the entire distributed system.
% Acquisition phase
The acquisition occurs during the \casePhaseExperimental{} \textbf{phase} because observability is primarily evaluated to assess and improve the design of monitoring and telemetry strategies rather than to monitor normal system operation.
% Acquisition method
The acquisition \textbf{method} is \caseAcquisitionExperimentalInstrumentation{} because estimating explainability and causal relationships requires advanced analysis techniques, such as causal inference or explainable AI methods, which are not provided by standard telemetry.

\subsection{Adaptivity Quotient} Adaptivity quotient is a metric designed to quantify how effectively and rapidly a DCC system can sense, respond to, and recover from changes or disruptions across the continuum (i.e., fluctuating workloads, node failures, shifting network conditions). As recent research introduces adaptive frameworks, the ability to adapt is crucial for ensuring service continuity, meeting application requirements, and optimizing resource use across the continuum. This metric is mathematically defined as shown in Eq.(\ref{eq_AQ}).
\begin{equation}\label{eq_AQ}
\mathcal{Q} = \frac{1}{K} \sum_{k=1}^{K} \left( \frac{P^{\text{post}}_k}{P^{\text{base}}_k} \cdot \frac{1}{T^{\text{adapt}}_k} \right)
\end{equation}
where, $K$ represents the total number of adaptation events observed in DCC. $P_k^{\text{before}}$ is the  metric (such as throughput or latency) before adaptation event $k$, and $P_k^{\text{after}}$ is the  metric after the system has adapted. $T_k$ is the time taken to complete the adaptation for event $k$. The minimal acquisition \textbf{scope} is \caseScopeSystem{} because adaptivity emerges from the coordinated behavior of the entire DCCS before, during, and after adaptation events.
% Acquisition phase
The acquisition occurs during the \casePhaseExperimental{} \textbf{phase} because its purpose is to compare different adaptation strategies, or configurations, under controlled conditions in order to identify the most effective design.
% Acquisition method
The acquisition \textbf{method} depends on the performance metric used to evaluate the adaptation. Metrics such as throughput, latency, or resource utilization can be obtained through \caseAcquisitionStandardTelemetry{}*, whereas application-specific or derived performance indicators may require \caseAcquisitionCustomInstrumentation{} or even \caseAcquisitionExperimentalInstrumentation{}.

\subsection{Maintain Equilibrium}\label{subsec:Equilibrium} 

More hardware means better performance, but also higher bills, greater energy use (on track to consume 7\% of global power by 2030~\cite{chen2025data}), and a larger environmental footprint. Every gain in QoS through replication, caching, or indexing carries a corresponding rise in cost and resource consumption; cut back, and reliability drops. This creates a feedback loop that is hard to escape. For instance, increasing replication factors to improve data availability demands additional storage and processing power, while low-latency techniques like indexing and caching require more computational resources. Conversely, reducing computational power or storage capacity affects QoS which creates a self-perpetuating challenge, such as more resources improve performance but increase costs. On the other side, cost-cutting compromises system reliability. These investments raise the question of how effectively increased resources translate into real-world performance gains. Notably, the growth of \textbf{personalized LLMs} may not require such large computing environments but necessitates careful control over infrastructure due to the associated costs, especially for small organizations (e.g., universities or public services). For example, Amdahl’s Law (Eq.~(\ref{eqAmdhals}), where the variable $S$ represents speedup and $F$ represents fraction), which predicted this situation \cite{amdahl2013computer}, highlights the limitation of diminishing returns: performance gains are constrained by the fraction of the workload that remains unoptimized. Therefore, the challenge for large-scale AI infrastructure initiatives lies not just in scaling hardware, but in efficiently using available resources to maximize QoS.
% \begin{equation}\label{eqAmdhals}
%     S_{overall} = {\left(1-F_{enhanced} + \frac{F_{enhanced}}{S_{enhanced}}\right)^{-1}}
% \end{equation}

The interplay between Cost ($Cost$), Resources ($R$), and QoS ($Q$) creates a complex optimization challenge. Increasing resources enhances QoS but raises costs, while reducing costs can degrade performance. Eq.~(\ref{eq_RCQ}) formalizes this trade-off, aiming QoS while adhering to cost and resource constraints.
\begin{equation}\label{eq_RCQ}
    \max_{R,Cost} Q(R,Cost) \text{ subjected to: } Cost\leq Cost_{max} \text{ , } R \geq R_{min}
\end{equation}
Current management approaches (i.e., predominantly static provisioning or reactive autoscaling) struggle to maintain this balance under volatile workloads. Closing the gap will require AI-driven, self-optimising frameworks that learn from ongoing system fluctuations and continuously steer the infrastructure toward a near-optimal equilibrium
across all three dimensions of Eq.~(\ref{eq_RCQ}). 

The minimal acquisition \textbf{scope} is \caseScopeSystem{} because equilibrium depends on the interaction between all resources and services across the DCCS.
% Acquisition phase
The acquisition occurs during the \casePhaseExperimental{} \textbf{phase} because the metric is intended to compare different architectures, orchestration strategies, or resource management policies under controlled workloads in order to identify the best trade-off between QoS, cost, and resource utilization.
% Acquisition method$
The acquisition \textbf{method} is \caseAcquisitionExperimentalInstrumentation{} because it requires combining heterogeneous metrics (e.g., QoS, cost, and resource usage) into an optimization model that is not directly available through standard telemetry.

\subsection{Data Locality Index}
In DCCS, tasks should ideally be executed close to the data source to reduce latency, bandwidth consumption, and privacy risks. A data locality index measures how close the execution node is to the data-producing node. 
Let $src(T_i)$ denote the source node that generates the data for task $T_i$, and let $exec(T_i)$ denote the node where the task is executed. The data locality index can be defined as:
\begin{equation}
    DLI = 1 - \frac{1}{|\mathcal{T}|}
    \sum_{T_i \in \mathcal{T}}
    \frac{dist(src(T_i), exec(T_i))}
    {dist_{max}},
\end{equation}
where $dist(src(T_i), exec(T_i))$ denotes the network, geographical, or hop-distance between the source and execution node, and $dist_{max}$ is the maximum possible distance in $\mathcal{S}$. A higher $DLI$ indicates better locality-aware execution. This metric is important because computing continuum applications often aim to reduce cloud dependency by processing data closer to IoT, mobile, edge, and fog nodes.  The minimal acquisition \textbf{scope} is \caseScopeMulti{} because computing the locality of a task requires information from both the data source node and the execution node.
% Acquisition phase
The acquisition occurs during the \casePhaseOperational{} \textbf{phase} in order to support locality-aware scheduling and task placement decisions that reduce communication latency, bandwidth consumption, and unnecessary data transfers.
% Acquisition method
The acquisition \textbf{method} is \caseAcquisitionStandardTelemetry{} because task placement, node identities, and network topology or routing information are routinely available through operating systems, orchestrators, and standard monitoring frameworks.

\subsection{Continuum Fragmentation Index}

As discussed before, DCC resources are distributed across cloud, fog, edge, mobile, and IoT layers. Even when the overall amount of free resources is sufficient, these resources may be fragmented across many nodes. This fragmentation can make it difficult to place large tasks, deploy service chains, or allocate workloads that require a minimum amount of capacity on the same node or within the same layer.

Let $R^{free}_j$ denote the free resource capacity of node $j$, and let
$R^{total}_{free}=\sum_{j\in\mathcal{N}}R^{free}_j$ represent the total free resource capacity across all nodes in the continuum. The Continuum Fragmentation Index can be defined as
\begin{equation}
    CFI =
    1 -
    \frac{\max_{j\in\mathcal{N}} R^{free}_j}
    {R^{total}_{free}}.
\end{equation}

A higher $CFI$ indicates higher fragmentation. This means that free resources exist in the DCC, but they are scattered across multiple nodes and may be difficult to use for placing large tasks or services. A lower $CFI$ indicates that the available resources are more consolidated, making task placement easier. When $CFI=0$, all free resources are concentrated in a single node. As $CFI$ approaches 1, the free resources become increasingly distributed across many nodes.  The minimal acquisition \textbf{scope} is \caseScopeSystem{} because fragmentation depends on the distribution of free resources across all participating nodes.
% Acquisition phase
The acquisition occurs during the \casePhaseOperational{} \textbf{phase} in order to identify resource fragmentation that may affect task placement during system execution.
% Acquisition method
The acquisition \textbf{method} is \caseAcquisitionStandardTelemetry{} because the required resource availability metrics are routinely exposed by OS, virtualization platforms, and orchestration frameworks.

\subsection{Migration Stability Index}
In DCCS, tasks and services may migrate among cloud, fog, edge, and mobile nodes. However, excessive migration can increase overhead, instability, and service disruption. A migration stability index measures whether the system adapts without unnecessary movement.
Let $M_k$ be the number of migrations during adaptation event $k$, and let $M^{max}_k$ be the maximum acceptable number of migrations. The migration stability index can be defined as:
\begin{equation}
    MSI =
    1 -
    \frac{1}{K}
    \sum_{k=1}^{K}
    \frac{M_k}{M^{max}_k}.
\end{equation}

A higher $MSI$ indicates that the system maintains stability with fewer migrations, while a lower value indicates migration-heavy adaptation. The minimal acquisition \textbf{scope} is \caseScopeSystem{} because migration stability depends on the migration activity across the entire DCCS.
% Acquisition phase
The acquisition occurs during the \casePhaseOperational{} \textbf{phase} in order to detect excessive migration, and support more stable orchestration decisions during system execution.
% Acquisition method
The acquisition \textbf{method} is \caseAcquisitionStandardTelemetry{} because migration events are routinely recorded by virtualization platforms, container orchestrators, and resource management frameworks.

\subsection{Adaptation Cost Efficiency}

Adaptation improves performance, but it also consumes resources, energy, and time. Therefore, it is useful to measure whether adaptation is worth its cost.

For an adaptation event $k$, let $\Delta P_k$ denote the performance improvement after adaptation, and let $Cost^{adapt}_k$ denote the total adaptation cost. The adaptation cost efficiency can be defined as:
\begin{equation}
    ACE =
    \frac{1}{K}\left(
    \sum_{k=1}^{K}
    \frac{\Delta P_k}
    {Cost^{adapt}_k}\right).
\end{equation}

The adaptation cost can include migration cost, energy cost, communication cost, and service interruption cost:
\begin{equation}
    Cost^{adapt}_k =
    Cost^{mig}_k +
    Cost^{eng}_k +
    Cost^{comm}_k +
    Cost^{down}_k.
\end{equation}

A higher $ACE$ indicates that the system achieves larger performance improvement with lower adaptation overhead. The minimal acquisition \textbf{scope} is \caseScopeMulti{} because both the adaptation benefits and their associated costs may involve multiple resources and services across the DCCS.
% Acquisition phase
The acquisition occurs during the \casePhaseExperimental{} \textbf{phase} because the metric is intended to compare different adaptation mechanisms or system configurations in order to identify the most cost-effective design under controlled conditions.
% Acquisition method
The acquisition \textbf{method} is \caseAcquisitionExperimentalInstrumentation{} because it combines multiple heterogeneous metrics, such as migration, energy, communication, and service interruption costs, into a derived optimization metric that is not directly available through standard telemetry.

\subsection{SLO Violation Recovery Time}

In DCCS, avoiding SLO violation is important, but recovering quickly after a violation is equally important. The SLO violation recovery time measures how long the system takes to return to an acceptable state after an SLO violation.

Let $t^{viol}_k$ be the time when SLO violation $k$ starts, and let $t^{rec}_k$ be the time when the system returns to the acceptable SLO range. The average SLO recovery time is:
\begin{equation}
    SRT =
    \frac{1}{K}
    \sum_{k=1}^{K}
    \left(t^{rec}_k - t^{viol}_k\right).
\end{equation}

A lower $SRT$ indicates faster recovery and stronger resilience. The minimal acquisition \textbf{scope} is \caseScopeApp{} because SLOs are typically defined for applications, although system-level SLOs can also be specified.
% Acquisition phase
The acquisition occurs during the \casePhaseOperational{} \textbf{phase} because the metric enables SLO-aware applications and autonomic management frameworks to detect violations and evaluate how quickly the application returns to an acceptable operating state.
% Acquisition method
The acquisition \textbf{method} is \caseAcquisitionExperimentalInstrumentation{} because, as discussed for the SLO compliance metric, it depends on an external SLO management component capable of defining SLOs, detecting violations, and identifying the recovery instant in addition to standard telemetry.

\subsection{Carbon Efficiency per Useful Task}
A more application-aware metric is carbon efficiency ($CO_2$ emissions) per useful task, which measures how many useful tasks are completed per unit carbon emission. 

Let $CO_2(\mathcal{S})$ denote the total carbon emission of the DCCS during $\Delta t$. The carbon efficiency can be defined as:
\begin{equation}
    CEUT =
    \frac{|\mathcal{T}^{s}|}
    {CO_2(\mathcal{S})}.
\end{equation}

For quality-aware applications, the metric can be extended as:
\begin{equation}
    QCEUT =
    \frac{\sum_{T_i\in\mathcal{T}^{s}} Q_i}
    {CO_2(\mathcal{S})}.
\end{equation}

A higher value means that the system produces more useful service output per unit carbon emission. The minimal acquisition \textbf{scope} is \caseScopeApp{} because, although both useful tasks and CO$_2$ emissions could be measured for a single computational node, this metric evaluates the useful work produced by an application as a whole relative to its carbon footprint.
% Acquisition phase
The acquisition occurs during the \casePhaseExperimental{} \textbf{phase} because the metric is primarily intended to compare the sustainability of different application designs or deployment strategies.
% Acquisition method
The acquisition \textbf{method} is \caseAcquisitionExperimentalInstrumentation{} because completed tasks require \caseAcquisitionCustomInstrumentation{}, while CO$_2$ emissions require \caseAcquisitionExperimentalInstrumentation{}. Consequently, the overall metric inherits the experimental acquisition method.

\subsection{Water Usage Efficiency}

For large-scale DCCS involving cloud and data-center resources, water consumption is becoming an important sustainability metric because data centers use water for cooling. Water usage efficiency can be defined as:
\begin{equation}
    WUE =
    \frac{W_{\mathcal{S}}}
    {E_{\mathcal{S}}},
\end{equation}
where $W_{\mathcal{S}}$ is the total water consumed by the infrastructure and $E_{\mathcal{S}}$ is the total energy consumed. 

A task-aware version can be defined as:
\begin{equation}
    WPT =
    \frac{W_{\mathcal{S}}}
    {|\mathcal{T}^{s}|},
\end{equation}
where $WPT$ denotes water consumption per successful task. The minimal acquisition \textbf{scope} is \caseScopeApp{} because, although energy consumption and successful tasks can be measured for a single computational node, this metric evaluates the water efficiency of an application as a whole.
% Acquisition phase
The acquisition occurs during the \casePhaseExperimental{} \textbf{phase} because the metric is primarily intended to compare the sustainability of different application designs, deployment strategies, or execution environments.
% Acquisition method
The acquisition \textbf{method} is \caseAcquisitionExperimentalInstrumentation{} because water consumption is generally not exposed by cloud providers or computing nodes through standard telemetry. Estimating the water footprint therefore requires provider-specific information or external estimation models, making novel instrumentation necessary.

\subsection{Data Freshness or Age of Information}

In DCCS, especially for IoT, cyber-physical systems, digital twins, and real-time monitoring, it is not enough that data arrives; it must arrive while still fresh. Age of Information measures the freshness of received data or how up-to-date the information available in a DCCS is with respect to when it was originally generated.

Let $u_i(t)$ denote the generation time of the most recently received update from source $i$ at time $t$. The age of information is:
\begin{equation}
    AoI_i(t) = t - u_i(t).
\end{equation}

The average age of information is:
\begin{equation}
    \overline{AoI} =
    \frac{1}{\Delta t}
    \int_{t_0}^{t_1}
    AoI_i(t)dt.
\end{equation}

For DCCS, system-level freshness can be defined as:
\begin{equation}
    F_{\mathcal{S}} =
    1 -
    \frac{1}{\rho}
    \sum_{i=1}^{\rho}
    \frac{\overline{AoI}_i}
    {AoI^{max}_i}.
\end{equation}

A higher freshness score indicates that the system maintains more up-to-date information from data-producing nodes. 
% Acquisition scope
The minimal acquisition \textbf{scope} is \caseScopeMulti{} because computing the age of information requires timestamps from both the data-producing node and the node that receives or processes the update.
% Acquisition phase
The acquisition occurs during the \casePhaseOperational{} \textbf{phase} in order to detect stale information and support freshness-aware scheduling, monitoring, or control decisions during system execution.
% Acquisition method
The acquisition \textbf{method} is \caseAcquisitionStandardTelemetry{} because data generation and reception timestamps are routinely available through distributed tracing frameworks.

\subsection{Trustworthiness Score}

In heterogeneous continuum environments, not all nodes may be equally trustworthy. Trustworthiness can combine reliability, security, privacy, availability, and historical behavior.

For node $j$, a trustworthiness score can be defined as:
\begin{equation}
    TS_j =
    \alpha_1 R_j
    +
    \alpha_2 A_j
    +
    \alpha_3 Sec_j
    +
    \alpha_4 Priv_j
    +
    \alpha_5 Hist_j,
\end{equation}
where $R_j$ is reliability, $A_j$ is availability, $Sec_j$ is security score, $Priv_j$ is privacy score, and $Hist_j$ is historical behavior score. The weights satisfy: $ \sum_{r=1}^{5}\alpha_r=1$.

The system-level trustworthiness score is:
\begin{equation}
    TS_{\mathcal{S}} =
    \frac{1}{\zeta}
    \sum_{j\in\mathcal{N}} TS_j.
\end{equation}
The minimal acquisition \textbf{scope} is \caseScopeNode{} because the trustworthiness score can be computed independently for each computational node, even though system-level trustworthiness can later be obtained by aggregation.
% Acquisition phase
The acquisition occurs during the \casePhaseOperational{} \textbf{phase} in order to support trust-aware scheduling and runtime decisions, allowing orchestration frameworks to avoid nodes with low trustworthiness.
% Acquisition method
The acquisition \textbf{method} is \caseAcquisitionExperimentalInstrumentation{} because several of its components, such as security, privacy, or historical behavior scores, are not directly exposed through standard telemetry and typically require dedicated assessment mechanisms.

\section{Conclusion}\label{sec:Conclusion}
Performance evaluation of DCCS requires a structured view of metrics because these systems combine heterogeneous computing, networking, storage, and sensing resources across cloud, fog, edge, mobile, IoT, and sensor layers. In this paper, we presented a taxonomy of performance metrics for DCCS and organized them into computing-level, network-level, and application/user-level categories. Instead of treating these metrics as independent measures, the taxonomy highlights how they collectively describe different aspects of continuum behavior, from task execution and resource usage to communication quality and application satisfaction. We also provided mathematical formulations for representative metrics to support clearer interpretation and more consistent comparison across different architectures, algorithms, and deployment scenarios. In addition, we discussed metric acquisition requirements in terms of acquisition scope, acquisition phase, and measurement method, which helps determine whether a metric should be collected from a single node, multiple nodes, or the full system, and whether it is better suited for operational monitoring or experimental evaluation. The paper further highlighted emerging dimensions such as sustainability, observability, adaptability, data locality, migration awareness, and continuum fragmentation, which reflect the evolving requirements of modern DCCS. The proposed taxonomy can serve as a practical reference for selecting suitable metrics according to the evaluation objective, application requirements, and deployment context. In future work, we will design computing continuum simulator that implement these metrics and use them to evaluate DCCS architectures and algorithms under realistic workload, mobility, resource, and network conditions.

\section*{Acknowledgment}
This research is partially funded by \textit{the European Union TALENTS (101299722)}, \textit{the Junta de Andalucía (Consejería de Universidad, Investigación e Innovación)} and co-funded by the EU through the \textit{FEDER Andalucía 2021-2027 Programme} under project \textit{DGP\_PIDI\_2024\_00918}, and also supported by CNS2023-144359 financed by MICIU/AEI/10.13039/501100011033 and the European Union NextGenerationEU/PRTR.

\bibliography{references}
\balance
%%%%%%%%%%%%%%%%%%%%%%% End of the Document
\end{document}